\documentclass[aps,pra,reprint]{revtex4-1}
\usepackage{amsmath}
\usepackage{resizegather}
\DeclareMathOperator{\acosh}{acosh}

\usepackage{amssymb}
\newcommand{\tr}{\mathsf{tr}}
\usepackage{braket}
\usepackage{dsfont}
\usepackage{mathtools}

\usepackage{epstopdf}
\usepackage{float}

\usepackage{tikz} 
\usetikzlibrary{backgrounds}
\usetikzlibrary{decorations.markings}

\usepackage{todonotes}
\usepackage{hyperref}
\newcommand{\Esq}{E_{\mathrm{sq}}}
\newcommand{\N}{\mathcal N}
\newcommand{\Dp}{\mathcal D_p}
\newcommand{\I}{\mathcal I}
\newcommand{\mms}{\pi}

\begin{document}
\title{Assessing the performance of quantum repeaters for all phase-insensitive Gaussian bosonic channels}
\author{K. Goodenough}
\email{kdgoodenough@gmail.com}
\author{D. Elkouss}
\author{S. Wehner}
\affiliation{QuTech, Delft University of Technology, Lorentzweg 1, 2628 CJ Delft, The Netherlands}

\begin{abstract}
One of the most sought-after goals in experimental quantum communication is the implementation of a quantum repeater. The performance of quantum repeaters can be assessed by comparing the attained rate with the quantum and private capacity of direct transmission, assisted by unlimited classical two-way communication. However, these quantities are hard to compute, motivating the search for upper bounds. Takeoka, Guha and Wilde found the squashed entanglement of a quantum channel to be an upper bound on both these capacities. In general it is still hard to find the exact value of the squashed entanglement of a quantum channel, but clever sub-optimal squashing channels allow one to upper bound this quantity, and thus also the corresponding capacities. 
Here, we exploit this idea to obtain bounds for any phase-insensitive Gaussian bosonic channel. This bound allows one to benchmark the implementation of quantum repeaters for a large class of channels used to model communication across fibers. In particular, our bound is applicable to the realistic scenario when there is a restriction on the mean photon number on the input.
Furthermore, we show that the squashed entanglement of a channel is convex in the set of channels, and we use a connection between the squashed entanglement of a quantum channel and its entanglement assisted classical capacity. Building on this connection, we obtain the exact squashed entanglement and two-way assisted capacities of the $d$-dimensional erasure channel and bounds on the amplitude-damping channel and all qubit Pauli channels. In particular, our bound improves on the previous best known squashed entanglement upper bound of the depolarizing channel.
\end{abstract}    
    \maketitle

\section{Introduction}
Optical quantum communication over long distances suffers from innate losses~\cite{jouguet2012field,shimizu2014performance,dixon2015high,pirandola2015high,korzh2015provably}. While in a classical setting the signal can be amplified at intermediate nodes to counteract this loss, this is prohibited in a quantum setting due to the no-cloning theorem~\cite{wootters1982single}. This problem can be overcome by implementing a quantum repeater, allowing entanglement over larger distances~\cite{briegel1998quantum,duan2001long}. The successful implementation of a quantum repeater will form an important milestone in the development of a quantum network \cite{perseguers2013distribution}. At this stage however, physical implementations perform worse than direct transmission \cite{sangouard2011quantum,yuan2008experimental}. 
As the experimental results improve it will be necessary to evaluate whether or not an implementation has achieved a rate not possible via direct communications. 
This can be done by comparing the attainable rate \emph{with} a quantum repeater~\cite{Bratzik_14,muralidharan2014ultrafast,Azuma_15,Krovi_15,Munro_15,Piparo_15b,luong2015overcoming,khalique2015practical} to the capacity of the associated quantum channel (i.e. direct transmission) for that task. For future quantum networks, arguably the two most relevant tasks are the transmission of quantum information and private classical communication. 
The capacity of a quantum channel for these two tasks, assuming that we allow the communicating parties to freely exchange classical communication, is given by the two-way assisted quantum and private capacity. We denote these quantities by $Q_2(\N)$ and $P_2(\N)$, respectively.

Finding exact values for $Q_2(\N)$ and $P_2(\N)$, however, is highly nontrivial thus motivating the search for upper bounds for them \cite{horodecki2005secure}. After having shown that the squashed entanglement of a channel is a quantity that is such an upper bound~\cite{takeoka2014squashed}, Takeoka, Guha and Wilde showed that there is a fundamental rate-loss trade-off in quantum key distribution and entanglement distillation over practical channels~\cite{takeoka2014fundamental}.

The squashed entanglement $\Esq(A;B)_{\rho}$ of a bipartite state $\rho_{AB}$ is a quantity defined as 
	\begin{gather}
	\label{eq:squashedentstate}
		\Esq(A;B)_{\rho} := \frac{1}{2}\inf\limits_{\mathcal{S}_{E\rightarrow E'}}I(A;B|E')\ ,
	\end{gather}
which was introduced by Christandl and Winter~\cite{christandl2004squashed} as an entanglement measure for a bipartite state. The squashed entanglement can be interpreted as the environment $E$ holding some purifying system of $\rho_{AB}$, and then squashing the correlations between $A$ and $B$ as much as possible by applying a channel $\mathcal{S}_{E\rightarrow E'}$ that minimizes the conditional mutual information $I(A;B|E')$. Extending this idea from states to channels, Takeoka, Guha and Wilde \cite{takeoka2014squashed,takeoka2014fundamental} defined the squashed entanglement $\Esq(\N)$ of a quantum channel as the maximum squashed entanglement that can be achieved between $A$ and $B$,
\begin{gather}
\label{eq:squashedentchannel}
\Esq(\N) := \max_{\Ket{\psi}_{AA'}} \Esq(A;B)_{\rho}\ ,
\end{gather}
where $\rho_{AB} = \N_{A'\rightarrow B}(\Ket{\psi}\Bra{\psi}_{AA'})$ is the state shared between Alice and Bob after the $A'$ system is sent through the channel $\N_{A'\rightarrow B}$.~They showed that $\Esq(\N)$ is an upper bound on the two two-way assisted capacities.

Unfortunately, there is no known algorithm for computing the squashed entanglement of a channel. This is partially due to the fact that the dimension of $E'$ is \emph{a priori} unbounded and that computing the squashed entanglement of a state is already an NP-hard problem~\cite{huang2014computing} and thus might even be uncomputable. However, fixing the channel in \eqref{eq:squashedentstate} in general yields an upper bound on $\Esq(\mathcal{N})$. Exploiting this idea of fixing a specific ``squashing channel'' $\mathcal{S}_{E\rightarrow E'}$, Takeoka et al. derived upper bounds on the squashed entanglement of several channels. Notably, they used this technique to find an upper bound for the pure-loss bosonic channel. 

The main contribution of this paper is an upper bound applicable to all phase-insensitive Gaussian bosonic channels. We apply this bound to the pure-loss channel, the additive noise channel and the thermal channel. 

Additionally, we obtain results for finite-dimensional channels by using tools that we develop here. The first of these consists of a concrete squashing channel that we call the trivial squashing channel which can be connected with the entanglement-assisted capacity. This connection, first observed by Takeoka et al. (see~\cite{bennett2014quantum}), allows us to compute the exact two-way assisted capacities of the $d$-dimensional erasure channel, and bounds on the amplitude damping channel and general Pauli channels.
Second, the squashed entanglement of entanglement breaking channels is zero. Third, for channels that can be written as a convex sum of channels the convex sum of the squashed entanglement of each channel is an upper bound, i.e.~$\Esq(\N)$ is convex on the set of channels. We combine all three of these tools to obtain bounds for the qubit depolarizing channel.
    
\section{Notation}
\noindent In this section we lay out the notation and conventions that we follow in this paper.

For a quantum state $\rho_{A}$ the von Neumann entropy of $\rho_{A}$ is defined as $H(A)=-\tr \rho_A\log\rho_A$. For convenience we take all logarithms in base two and set $\log_2(\cdot) \equiv \log(\cdot)$. 
For a quantum state $\rho:=\rho_{AB}$ the conditional entropy of system $A$ given $B$ is defined as $H(A|B)_{\rho}=H(AB){\rho}-H(B)_{\rho}$. Here $H(B)$ is computed over the state $\rho_B=\tr_A(\rho_{AB})$, where we denote the partial trace over system $A$ of a state $\rho_{AB}$ by $\tr_A(\rho_{AB})$. For a tripartite state $\rho_{ABE}$ the conditional mutual information is defined as $I(A;B|E) = H(A|E)-H(A|BE)$. Whenever there is confusion regarding the state over which we are computing an entropic quantity we will add the state as a subscript.

A quantum channel $\N_{A'\rightarrow B}$ is a completely positive and trace preserving map~\cite{wilde2011classical} between linear operators on Hilbert spaces $\mathcal{H}_{A'}$ and $\mathcal{H}_B$. A quantum channel $\mathcal{N}$ can always be embedded into an isometry $V^\N_{A'\rightarrow BE}$ that takes the input to the output system $B$ together with an auxiliary system $E$ that we call the environment. This isometry is called the Stinespring dilation of the channel. The action of the channel is recovered by tracing out the environment: $\N(\rho)=\tr_E(V\rho V^*)$. 

We denote the $d$-dimensional maximally mixed state by $\mms$. The dimension of $\mms$ is implicit and should be clear from the context. Let $\mathcal{N}$ be a channel with input and output dimension $d$. Then $\mathcal{N}$ is unital if $\mathcal{N}(\mms) = \mms$. 

\section{Some properties of $\Esq(\N)$}
\noindent In this section we prove several properties of $\Esq(\N)$ that will be of general use for obtaining upper bounds on the squashed entanglement of concrete channels. First we define a squashing channel that we call the trivial squashing channel and connect it to the entanglement assisted capacity of that channel, an observation previously made in~\cite{bennett2014quantum} by Takeoka et al. Second, we prove that the squashed entanglement of entanglement breaking channels is zero. The third property is that $\Esq(\N)$ is convex in the set of channels.

\subsection{The trivial squashing channel}
\noindent One possible squashing channel $\mathcal{S}_{E\rightarrow E'}$ is the identity channel, which we will call the trivial squashing channel. The state on $ABE'$ is pure, from which it can easily be calculated that
\begin{align}
\Esq(\mathcal{N}) &\leq \max_{\ket{\phi}_{AA'}}\frac{1}{2}I(A;B|E)\\
                             &= \max_{\ket{\phi}_{AA'}}\frac{1}{2}(H(A|E)-H(A|BE))\\
                             &= \max_{\ket{\phi}_{AA'}}\frac{1}{2}(H(AE)-H(E)\nonumber \\
                             &-H(ABE)+H(BE))\\
                             &= \max_{\ket{\phi}_{AA'}}\frac{1}{2}(H(B)+H(A)-H(AB))\\                      
                             &= \max_{\ket{\phi}_{AA'}}\frac{1}{2}I(A;B)\ .
\label{eq:ce}
\end{align}
The maximization in the right hand of \eqref{eq:ce}, up to the $1/2$ factor, characterizes the capacity of a quantum channel for transmitting classical information assisted by unlimited entanglement~\cite{bennett1999entanglement}. In other words, the squashed entanglement is bounded from above by one half the entanglement assisted capacity of the channel which we denote by $C_E(\N)$. This connection, which was first observed by Takeoka et al. (see~\cite{bennett2014quantum}), allows us to bound the squashed entanglement for all channels for which $C_E(\N)$ is known. 

\subsection{Entanglement breaking channels}
\noindent Entanglement breaking channels have zero private and quantum capacities assisted by two-way communications. We show that the squashed entanglement of these channels is also zero, following a similar approach as was done for the squashed entanglement of separable states in \cite{thesischristandl}. In order to see this note that if an entanglement breaking channel $\N_{\mathrm{EB}}$ is applied to half of a bipartite state, the output is always separable and can be written as a convex combination of product states,
\begin{align}
\psi_{AB}&=\I\otimes\N_{\mathrm{EB}}(\ket{\psi}\bra{\psi}_{AA'})\\
               &=\sum_i\lambda_i\ket{\alpha_i}\bra{\alpha_i}_{A}\otimes\ket{\beta_i}\bra{\beta_i}_{B},
\end{align}
where we denote by $\I$ the identity map. A possible purification of $\psi_{AB}$ is
\begin{equation}
\ket{\psi}_{ABE_1E_2}= \sum_i\sqrt\lambda_i\ket{\alpha_i}_{A}\ket{\beta_i}_{B}\ket{i}_{E_1}\ket{i}_{E_2}\ ,
\end{equation}
where $\lbrace \Ket{i}_{E_1}\rbrace$ and $\lbrace \Ket{i}_{E_2}\rbrace$ are sets of orthonormal states. If the squashing channel consists of tracing out the $E_2$ system, the resulting state is

\begin{gather}
\sum_i \lambda_i \Ket{\alpha_i}\Bra{\alpha_i}_A \otimes \Ket{\beta_i}\Bra{\beta_i}_B \otimes \Ket{i}\Bra{i}_{E_1}\ ,
\end{gather}
which has zero conditional mutual information.

\subsection{Convexity of $\Esq(\N)$ in the set of channels}
\noindent The squashed entanglement of the channel is convex in the set of channels. We prove this in the Appendix following similar ideas to the ones used in \cite{christandl2004squashed} to prove that the squashed entanglement is convex in the set of states.
Hence, if $\N = \sum_j p_j \N_j$ with $\sum_j p_j = 1$ and $p_j\geq 0$, then

\begin{gather}
\Esq(\N) \leq \sum_j p_j \Esq(\N_j)\ .
\end{gather}

\section{Finite-dimensional channels}
\label{sec:finch}
\noindent To build intuition before moving to bosonic channels, let us first bound the squashed entanglement of finite-dimensional channels, i.e. channels where both the input and output dimensions are finite.

An illustrative example of the effectiveness of the trivial squashing channel is the $d$-dimensional erasure channel $\mathcal{E}^d_{p}(\rho) = (1-p)\rho + p \Ket{e}\Bra{e}$, where $\rho$ is a $d-$dimensional state and $\Ket{e}$ is an erasure flag orthogonal to the support of any $\rho$ on the input \cite{wilde2011classical}. It is well known that $C_E(\mathcal{E}^d_p) = 2(1-p)\log(d)$~\cite{wilde2011classical} and that $Q_2(\mathcal{E}^d_p) = (1-p)\log(d)$~\cite{bennett1997capacities}. In general we have 
\begin{equation}
Q_2(\mathcal{N}) \leq P_2(\mathcal{N})\leq \Esq(\mathcal{N})\leq \frac{1}{2}C_E(\mathcal{N})\ ,
\end{equation} 
where the first inequality holds since the squashed entanglement of a channel is an upper bound on $Q_2(\N)$ and the second inequality follows from applying the trivial squashing channel. In the specific case of the erasure channel, we then must have that 
\begin{equation}
Q_2(\mathcal{E}^d_p) = P_2(\mathcal{E}^d_p) = \Esq(\mathcal{E}^d_p) = (1-p)\log(d)\ .
\end{equation}
That is, the trivial squashing channel is the \emph{optimal} squashing channel, yielding both two-way assisted capacities and the squashed entanglement of the $d$-dimensional erasure channel. We note that, up until now, this class of channels is the only class whose squashed entanglement has been calculated exactly. Independently of our work, in~\cite{pirandola2015general} the two-way assisted capacities of the $d$-dimensional erasure channel are established by computing the entanglement flux of the channel, which is also an upper bound on $P_2$.

\begin{figure}
	\centering
\includegraphics[clip, trim=36mm 0mm 81mm 0mm, scale = 0.110]{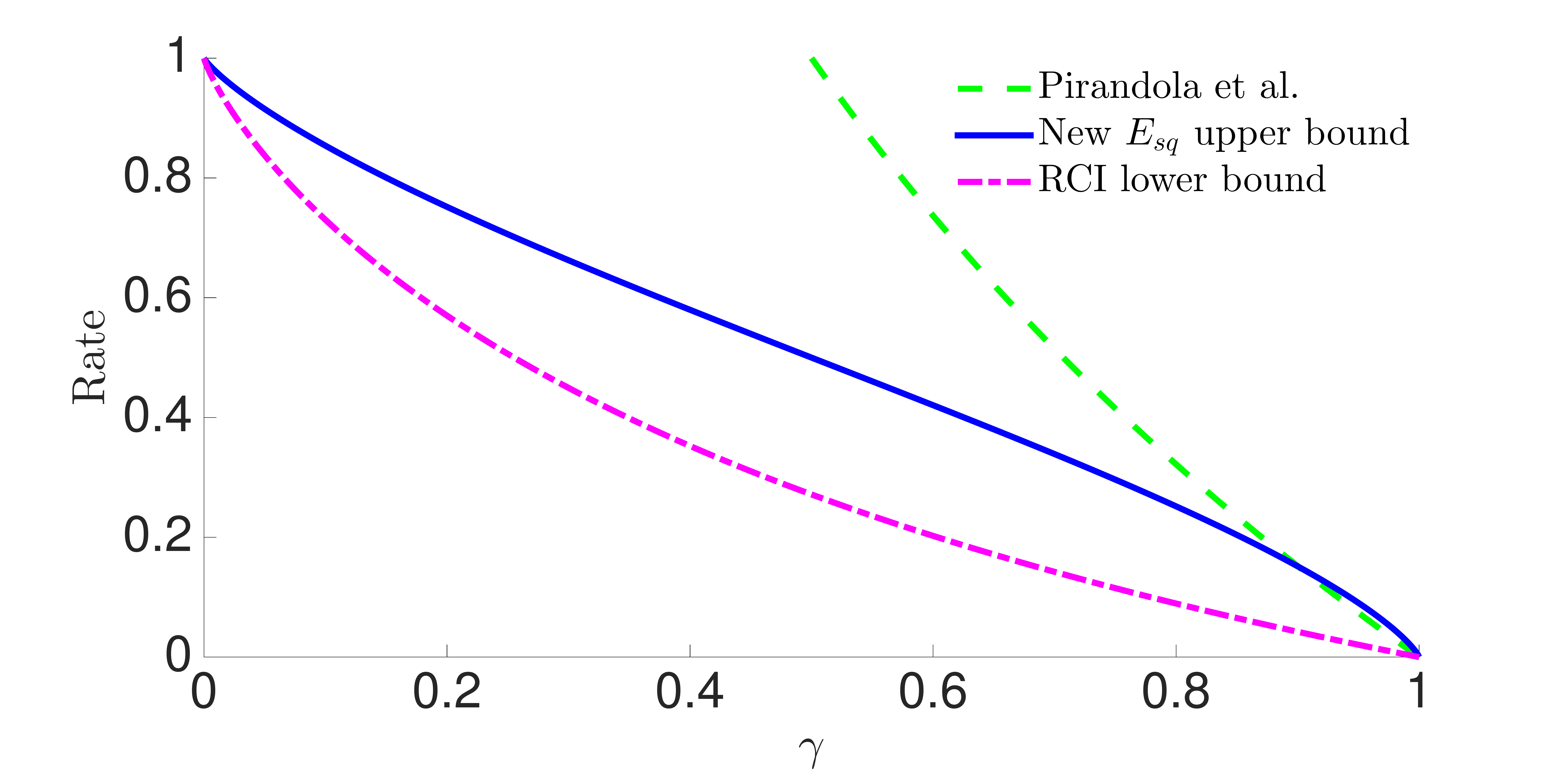}
\caption{Comparison of bounds for the amplitude damping channel. In dashed green the upper bound by Pirandola et al.~\cite{Pirandola:2015aa}, in solid blue the upper bound found in this paper and the dash-dotted magenta line is a lower bound given by the reverse coherent information~\cite{garcia2009reverse}.
}
\label{fig:amplitude}
\end{figure}

A second channel we can apply the trivial isometry to is the qubit damping channel $\N_{AD}^{\gamma}$, a channel that models energy dissipation in two-level systems. The qubit amplitude damping channel is defined as

\begin{align}
\N_{AD}^{\gamma}(\rho) := \sum_{i=0}^1A_i\rho A_i^{\dagger}\ ,
\end{align}
where
\begin{align}
A_0 = \begin{bmatrix}
1 & 0\\
0 & \sqrt{1-\gamma}
\end{bmatrix},~A_1 = \begin{bmatrix}
0 & \sqrt{\gamma}\\
0 & 0
\end{bmatrix}
\end{align}
with amplitude damping parameter $\gamma \in \lbrack 0, 1\rbrack$. Since the entanglement assisted classical capacity of the amplitude damping channel is known~\cite{wilde2011classical} to be equal to

\begin{align}
C_E\left(\N_{AD}^{\gamma}\right) = \max_{p\in \lbrace 0,1\rbrace}\lbrack h(p)+h((1-\gamma)p)-h(\gamma p)\rbrack\ ,
\end{align}
where $h(x) = -x\log(x)-(1-x)\log(1-x)$ is the binary entropy, we immediately find the bound

\begin{align}
P_2\left(\N_{AD}^{\gamma}\right) \leq \Esq\left(\N_{AD}^{\gamma}\right) \leq \frac{1}{2}C_E\left(\N_{AD}^{\gamma}\right)\ .
\end{align}
A comparison of this bound with the best known lower bound, given by the reverse coherent information (RCI) $\max_{p}\lbrack h(p)-h(p\gamma)\rbrack$, and an upper bound $P_2\left(\N_{AD}^{\gamma}\right) \leq \min\lbrace 1, -\log\gamma \rbrace$ found by Pirandola et al.~\cite{Pirandola:2015aa} using an entanglement flux approach, can be seen in Figure \ref{fig:amplitude}.

A third interesting example are $d$-dimensional unital channels for which the maximally entangled state on $AA'$ maximizes the mutual information $I(A;B)$. For these channels the trivial squashing channel gives the following compact upper bound
\begin{align}
\Esq(\mathcal{N}) &\leq \frac{1}{2}I(A;B) \\
&=\frac{1}{2}\lbrack H(A)+H(B)-H(AB)\rbrack\\
&= \log(d)-\frac{1}{2}H(E)\ .
\end{align}
In particular, this bound holds for any Pauli channel, where we have that $d = 2$. Any Pauli channel can be written as 
\begin{equation}
\mathcal{P}(\rho) = p_0\rho + p_1X\rho X+p_2XZ\rho ZX + p_3Z\rho Z\ ,
\end{equation}
with $\sum_{i=0}^3p_i = 1$. Choosing without loss of generality the maximally entangled state $\Ket{\Phi^+}_{AA'} = \frac{1}{\sqrt{2}}\lbrack \Ket{00}+\Ket{11} \rbrack_{AA'}$ as input on $AA'$, we see that the output has a purification of the form
\begin{align}
\sqrt{p_0}&\Ket{\Phi^+}_{AB}\Ket{00}_E+\sqrt{p_1}\Ket{\Psi^+}_{AB}\Ket{01}_E\nonumber\\
&+\sqrt{p_2}\Ket{\Psi^-}_{AB}\Ket{10}_E+\sqrt{p_3}\Ket{\Phi^-}_{AB}\Ket{11}_E\ .
\end{align}
From orthogonality of the Bell states, it can be seen that the entropy of the environment coincides with the classical entropy of the probability vector $\overline{p} = (p_0,p_1,p_2,p_3)$. That is, $H(E) = H(\overline{p})$ with $H(\overline{p})\equiv-\sum_{i=0}^3p_i\log p_i$. From this it follows that 
\begin{equation}
\label{eq:paulibound}
\Esq (\mathcal{P})\leq 1-\frac{1}{2}H(\overline{p})\ .
\end{equation}
Hence, we also obtain that $2-H(\overline{p})$ is the entanglement assisted classical capacity of a Pauli channel $\mathcal{P}$. 

\begin{figure}
	\centering
\includegraphics[clip, trim=36mm 0mm 81mm 0mm, scale = 0.110]{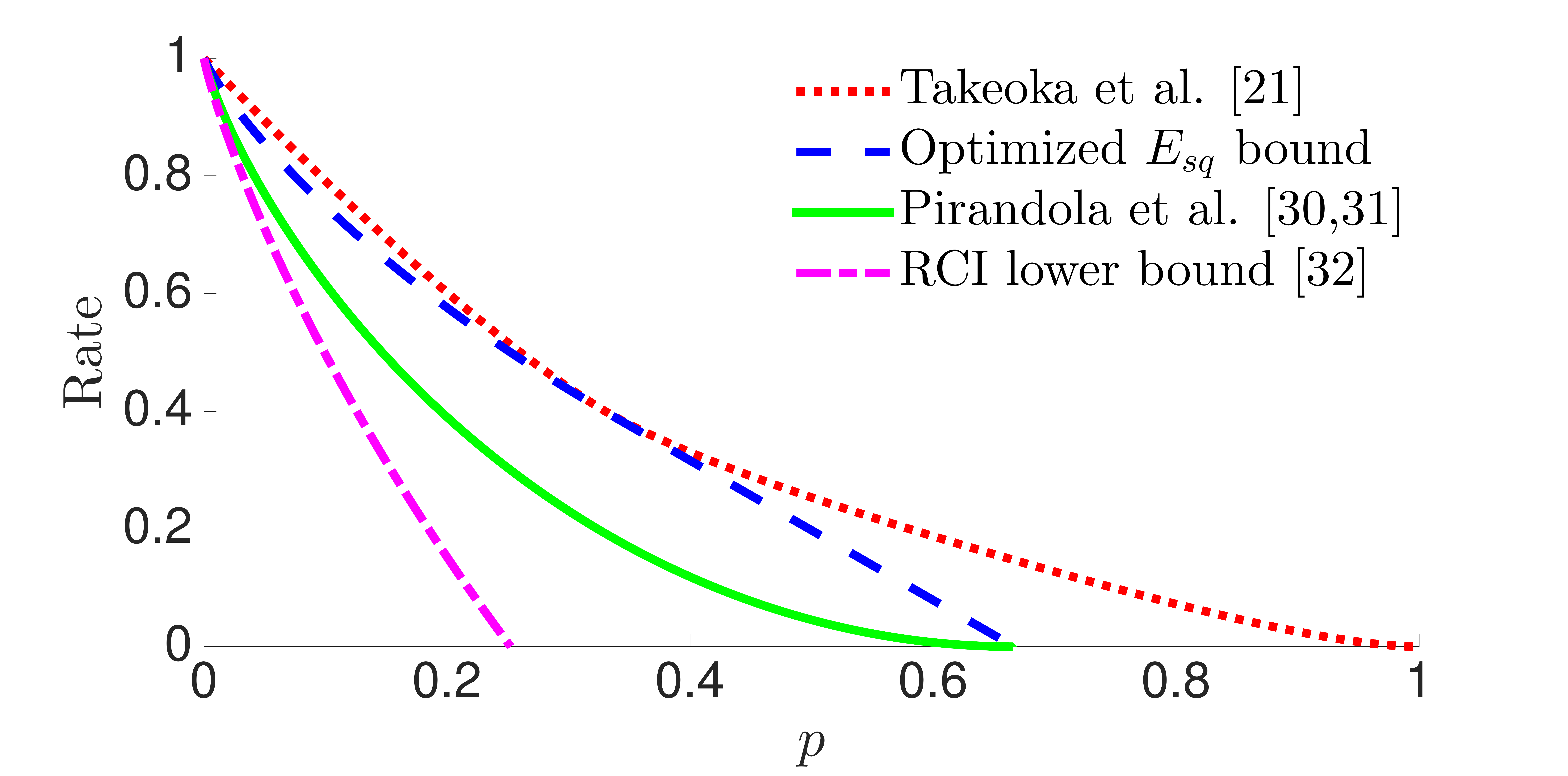}
\caption{Comparison of bounds for the depolarizing channel. The dotted red line is the upper bound by Takeoka et al.~\cite{takeoka2014squashed}, the dashed blue line is the optimized squashed entanglement bound in this paper, the solid green line is the entanglement flux upper bound by Pirandola et al.~\cite{pirandola2015general, Pirandola:2015aa} and the magenta line is a lower bound given by the reverse coherent information~\cite{garcia2009reverse}.
}
\label{fig:DepolarizingComparison}
\end{figure}

Let us now apply the bound for Pauli channels to a concrete channel, the (binary) depolarizing channel $\Dp$. The action of this channel is $\Dp(\rho)\equiv (1-p)\rho+p\mms$ for $p \in \lbrack0, 1 \rbrack$. This corresponds with the Pauli channel given by $\overline{p} = (1-\frac{3p}{4},\frac{p}{4},\frac{p}{4},\frac{p}{4})$. After this identification we find that 
\begin{gather}
\label{eq:dp1}
\Esq(\Dp)\leq \frac{3 p \log(p)+(4-3p)\log(4-3p)}{8}\ .
\end{gather}
The depolarizing channel can also be written as a convex combination of two other depolarizing channels, allowing us to use the convexity of $\Esq(\N)$ in the set of channels to improve on the upper bound in equation \eqref{eq:dp1}. We can compute the squashed entanglement of each individual channel and multiply it by the appropriate weight. Using this idea (see section~\ref{sec:depol} in the Appendix), we obtain the following stronger upper bound
\begin{equation}
\label{eq:dp2}
\Esq(\Dp)\leq \min_{0\leq\epsilon\leq p} (1-\alpha)\frac{3\epsilon\log(\epsilon)+(4-3\epsilon)\log (4-3\epsilon)}{8}\ .
\end{equation}
where $\alpha=\frac{p-\epsilon}{2/3-\epsilon}$. This bound is equal to \eqref{eq:dp1} for $0\leq p\lesssim \frac{1}{3}$, after which it linearly goes to zero at $p = \frac{2}{3}$. See Figure \ref{fig:DepolarizingComparison} for a comparison of this new bound, the bound by Takeoka et al.~\cite{takeoka2014squashed,takeoka2014squashed2}, the bound by Pirandola et al.~\cite{Pirandola:2015aa}, and the reverse coherent information~\cite{garcia2009reverse}.

\section{Phase-insensitive Gaussian bosonic channels}
\subsection{An upper bound on phase-insensitive channels}
\noindent In this section we discuss our main result, an upper bound on the squashed entanglement of any phase-insensitive Gaussian bosonic channel. Gaussian bosonic channels are of interest because they are used to model a large class of relevant operations on bosonic systems~\cite{Weedbrook:2012aa}. Phase-insensitive channels are those Gaussian bosonic channels which add equal noise in each quadrature of the bosonic systems. Imperfections in experimental setups for quantum communication with photons are modeled by phase-insensitive channels, motivating us to upper bound the squashed entanglement of all such channels. In particular this motivates the search for bounds where the input of the channel has a constraint on the mean photon number $N$.\\

\begin{figure}
\begin{centering}
\includegraphics{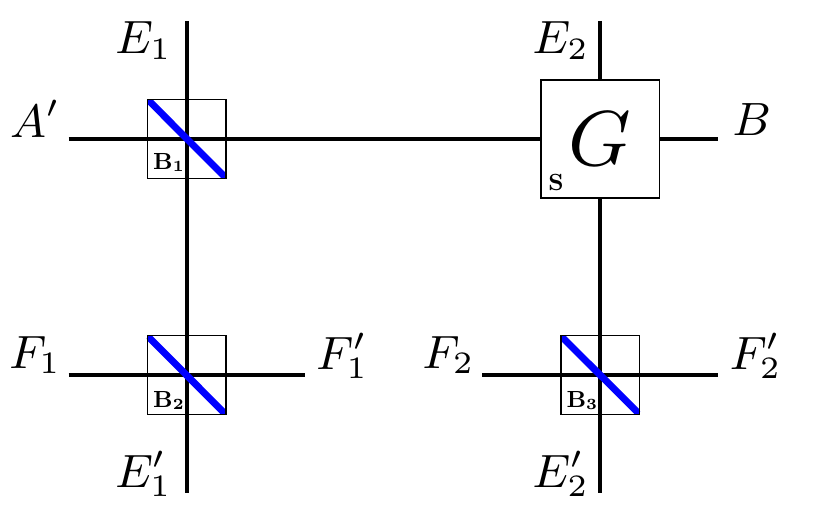}
\caption{A squashing isometry for any phase-insensitive Gaussian channel $\mathcal{N}_{\mathrm{PI}}$ taking $A'$ to $B$. The beamsplitter $\mathbf{B_1}$ and the two-mode squeezer $\mathbf{S}$ form the Stinespring dilation, while the balanced beamsplitters $\mathbf{B_2}$ and $\mathbf{B_3}$ form the squashing map. The beamsplitter $\mathbf{B}_1$ interacts with the vacuum on $E_1$ and $A$, and the two-mode squeezer $\mathbf{S}$ interacts with the output of $\mathbf{B}_2$ and the vacuum on $E_2$. The squashing isometry consists of two balanced beamsplitters $\mathbf{B_2}$ and $\mathbf{B_3}$ interacting with the vacuum on $F_1$ and $F_2$ and the output of the beamsplitter $\mathbf{B}_1$ and the two-mode squeezer $\mathbf{S}$.}
\label{fig:isometry}
\end{centering}
\end{figure}

\noindent Any phase-insensitive channel $\mathcal{N}_{\mathrm{PI}}$ is completely characterized by its a loss/gain parameter $\tau$ and noise parameter $\nu$. The Stinespring dilation of such a channel consists of a beamsplitter with transmissivity $T = \frac{2\tau}{\tau + \nu + 1}$ interacting with the vacuum on $E_1$, and a two-mode squeezer with squeezing parameter $r = \acosh(\sqrt{G})$ with the amplification $G = \frac{\tau + \nu + 1}{2} \geq 1$ interacting with the vacuum on $E_2$~\cite{garcia2012majorization} (see Figure \ref{fig:isometry} and the Appendix for a detailed definition of the channel). $T$ and $G$ also completely characterize any phase-insensitive channel. Takeoka et al.~\cite{takeoka2014squashed,takeoka2014squashed2,takeoka2014fundamental} found bounds for such channels by only considering the beamsplitter part of the Stinespring dilation. To be a valid channel, we must have that $\nu \geq \left|1-\tau\right|$. We further have that phase-insensitive channels are entanglement breaking whenever $\nu \geq \tau + 1$~\cite{giovannetti2014ultimate}, or equivalently, $G(1-T)\geq 1$. Hence, the squashed entanglement must be zero for channels with such parameters as discussed in the tools section.

\noindent Since we are interested in phase-insensitive Gaussian channels, we make the ansatz that a good squashing map will be a phase-insensitive channel. Numerical work suggests that, if only phase-insensitive isometries are considered, the pure-loss channel and the amplification channel separately have as optimal squashing isometry the balanced beamsplitter interacting with the vacuum. This motivates us to use the isometry consisting of two balanced beamsplitters at the outputs of the first beamsplitter and the two-mode squeezer (see Figure \ref{fig:isometry}). 
Using this isometry we obtain a bound for all phase-insensitive channels with restricted mean photon number $N$ (see Appendix for a derivation and a proof that the equation is monotonically non-decreasing as a function of $N$). This equation equals
\begin{gather}
g\left(\left(\nu_{BE_1'E_2'}\right)_{1}\right) + g\left(\left(\nu_{BE_1'E_2'}\right)_{2}\right) - g\left(\left(\nu_{E_1'E_2'}\right)_{1}\right) - g\left(\left(\nu_{E_1'E_2'}\right)_{2}\right)\label{eq:finiten2}\ ,
\end{gather}
with $g(x) = \left(\frac{x+1}{2}\right)\log(\frac{x+1}{2})-\left(\frac{x-1}{2}\right)\log(\frac{x-1}{2})$ \cite{Weedbrook:2012aa} and
\begin{align}
 \resizebox{0.477 \textwidth}{!} 
{
$\left(\nu_{E_1'E_2'}\right)_{1}~ = \left|\sqrt{-\frac{1+G^2+2N\left(1-T+GT\left(G-1\right)\right)+N^2\left(GT-1\right)^2+\left(G-1+N(GT-1)\right)\Omega^-}{2}}\right|$
}\nonumber \\
 \resizebox{0.475 \textwidth}{!} 
{
$\left(\nu_{E_1'E_2'}\right)_{2}~ = \left|\sqrt{-\frac{1+G^2+2N\left(1-T+GT\left(G-1\right)\right)+N^2\left(GT-1\right)^2-\left(G-1+N(GT-1)\right)\Omega^-}{2}}\right|$}\nonumber \\
 \resizebox{0.475 \textwidth}{!} 
{$\left(\nu_{BE_1'E_2'}\right)_{1} = \left|\sqrt{-\frac{1+G^2+2N\left(1-T+GT\left(G+1\right)\right)+N^2\left(1+GT\right)^2+\left(1+G+N(1+GT)\right)\Omega^+}{2}}\right|$}\nonumber \\
 \resizebox{0.475 \textwidth}{!} 
{
$\left(\nu_{BE_1'E_2'}\right)_{2} = \left|\sqrt{-\frac{1+G^2+2N\left(1-T+GT\left(G+1\right)\right)+N^2\left(1+GT\right)^2-\left(1+G+N(1+GT)\right)\Omega^+}{2}}\right|$}\nonumber
\end{align}
where we have set 
\begin{align}
 \resizebox{0.422 \textwidth}{!} 
{
$\Omega^{\pm} = \sqrt{(1+N)^2-4NT\pm 2G(1+N)(NT-1)+(G+GNT)^2}$}\ .
\end{align}
As $N\rightarrow \infty$, the bound above converges to its maximum value of
\begin{gather}
\Esq\left(\mathcal{N}_{\mathrm{PI}}\right) \leq \frac{\left(1-T^2\right)G\log(\frac{1+T}{1-T})-\left(G^2-1\right)T\log(\frac{G+1}{G-1})}{1-G^2T^2}\ ,
\label{eq:Gbound1}
\end{gather}
Rewriting the upper bound as function of the channel parameters $\tau$ and $\nu$~\cite{Weedbrook:2012aa} we obtain the upper bound
\begin{gather}
\Esq\left(\mathcal{N}_{\mathrm{PI}}\right) \leq \frac{\zeta(1+\nu+3\tau,1+\nu-\tau)-\tau \zeta(\tau+\nu+3,\tau+\nu-1)}{2(1+\nu+\tau)(1-\tau^2)}\ ,
\label{eq:Gbound2}
\end{gather}
where $\zeta(a,b) = ab\log(\frac{a}{b})$.

\subsection{Application to concrete phase-insensitive Gaussian channels with unconstrained photon input}
\subsubsection{Quantum-limited phase-insensitive channels}
\noindent A pure-loss channel has $G=1$. As a consequence, for pure-loss channels the bound in equation \eqref{eq:Gbound1} reduces to $\log(\frac{1+T}{1-T})$. This bound coincides with the bound found by Takeoka et al. 

\noindent In the opposite extreme we find quantum-limited amplifying channels, that is channels with $T = 1$ and $G>1$. For these channels, the bound by Takeoka is equal to infinity while \eqref{eq:Gbound1} is non-trivial. Concretely, it reduces to the finite value of $\log(\frac{G+1}{G-1})$. This should be compared with the exact capacities independently found by Pirandola et al.~\cite{pirandola2015general, pirandola2015ultimate, Pirandola:2015aa} using an entanglement flux approach, $Q_2 = P_2 = \log(\frac{G}{G-1})$.

\subsubsection{Additive noise channel}
\begin{figure}
	\centering
\includegraphics[clip, trim=59mm 0mm 75mm 0mm, scale = 0.10]{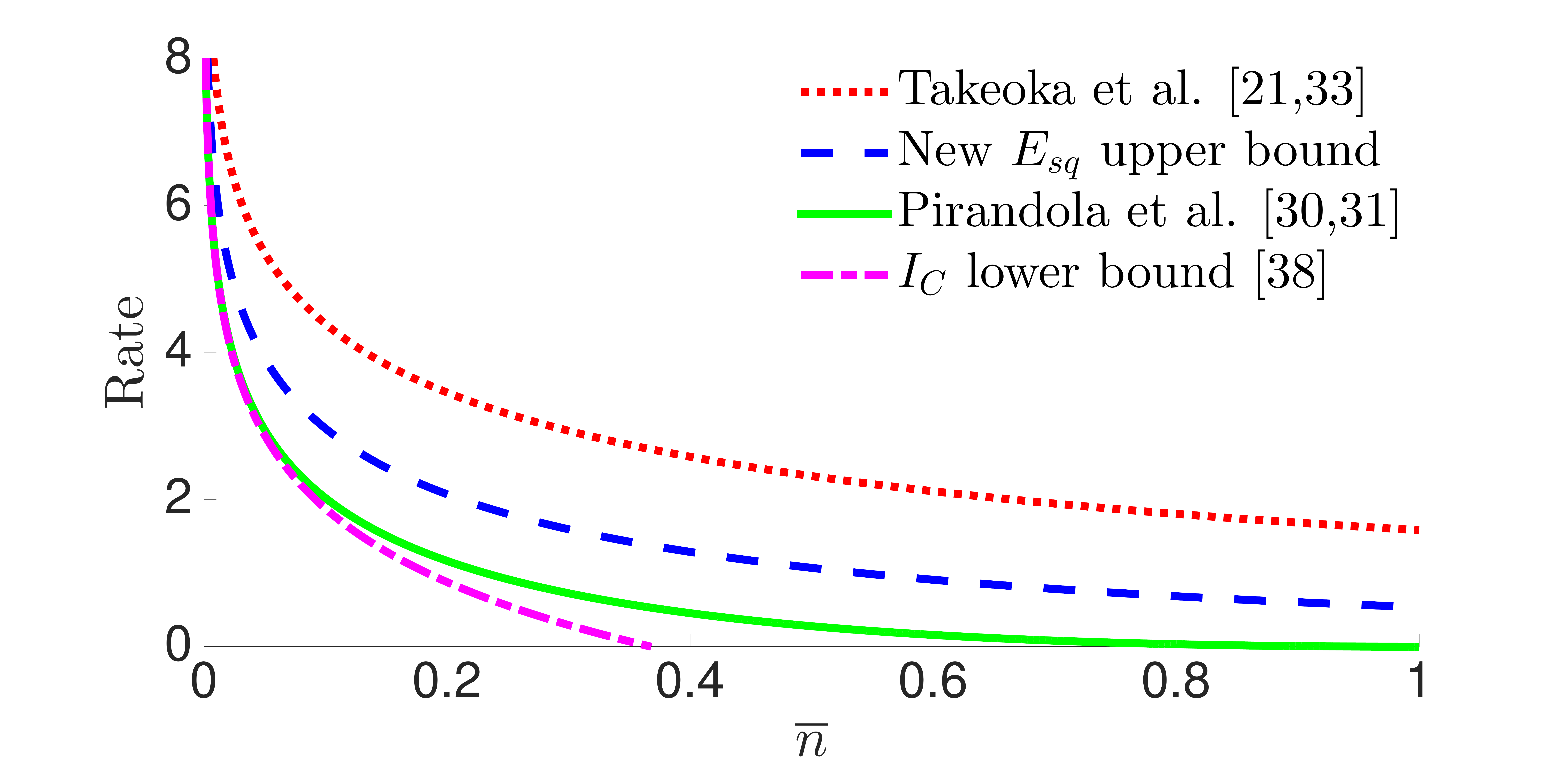}
\caption{Comparison of the upper bounds mentioned in this paper for the additive noise channel. The dotted red line is the upper bound by Takeoka et al.~\cite{takeoka2014squashed}, the dashed blue line is the squashed entanglement bound in this paper, the solid green line is the entanglement flux upper bound by Pirandola et al.~\cite{pirandola2015general, Pirandola:2015aa} and the magenta line is the coherent information of the channel which is a lower bound~\cite{holevo2001evaluating}.}
\label{fig:additive}
\end{figure}

\noindent An additive noise channel only adds noise to the input, without damping or amplifying the signal. For an additive noise channel $\mathcal{N}_{\mathrm{add}}$ we have $T = \frac{1}{\overline{n}+1}$ and $G = \frac{1}{T} = \overline{n}+1$, where $\overline{n}$ is the noise variance. Taking the limit of equation \eqref{eq:Gbound1} as $G\rightarrow \frac{1}{T} = \overline{n}+1$ we show in the Appendix that the upper bound becomes
\begin{align}
\Esq\left(\mathcal{N}_{\mathrm{add}}\right)& \leq \frac{T^2+1}{2T}\log(\frac{1+T}{1-T})-\frac{1}{\ln2}\\
&= \frac{\overline{n}^2+2\overline{n}+2}{2\overline{n}+2}\log(\frac{\overline{n}+2}{\overline{n}})-\frac{1}{\ln2}\ .
\end{align}
This should be compared with the upper bound independently found by Pirandola et al.~\cite{pirandola2015general, pirandola2015ultimate, Pirandola:2015aa}, $\frac{\overline{n}-1}{\ln(2)}-\log\overline{n}$ and the coherent information $I_C(\mathcal{N}_{\mathrm{add}}) = -\frac{1}{\ln(2)}-\log\overline{n}$ which is a lower bound on $P_2(\N)$~\cite{holevo2001evaluating}. See Figure \ref{fig:additive} for a comparison of these bounds.

\subsubsection{Thermal channel}
\begin{figure}
	\centering
\includegraphics[clip, trim=19mm 0mm 71mm 0mm, scale = 0.10]{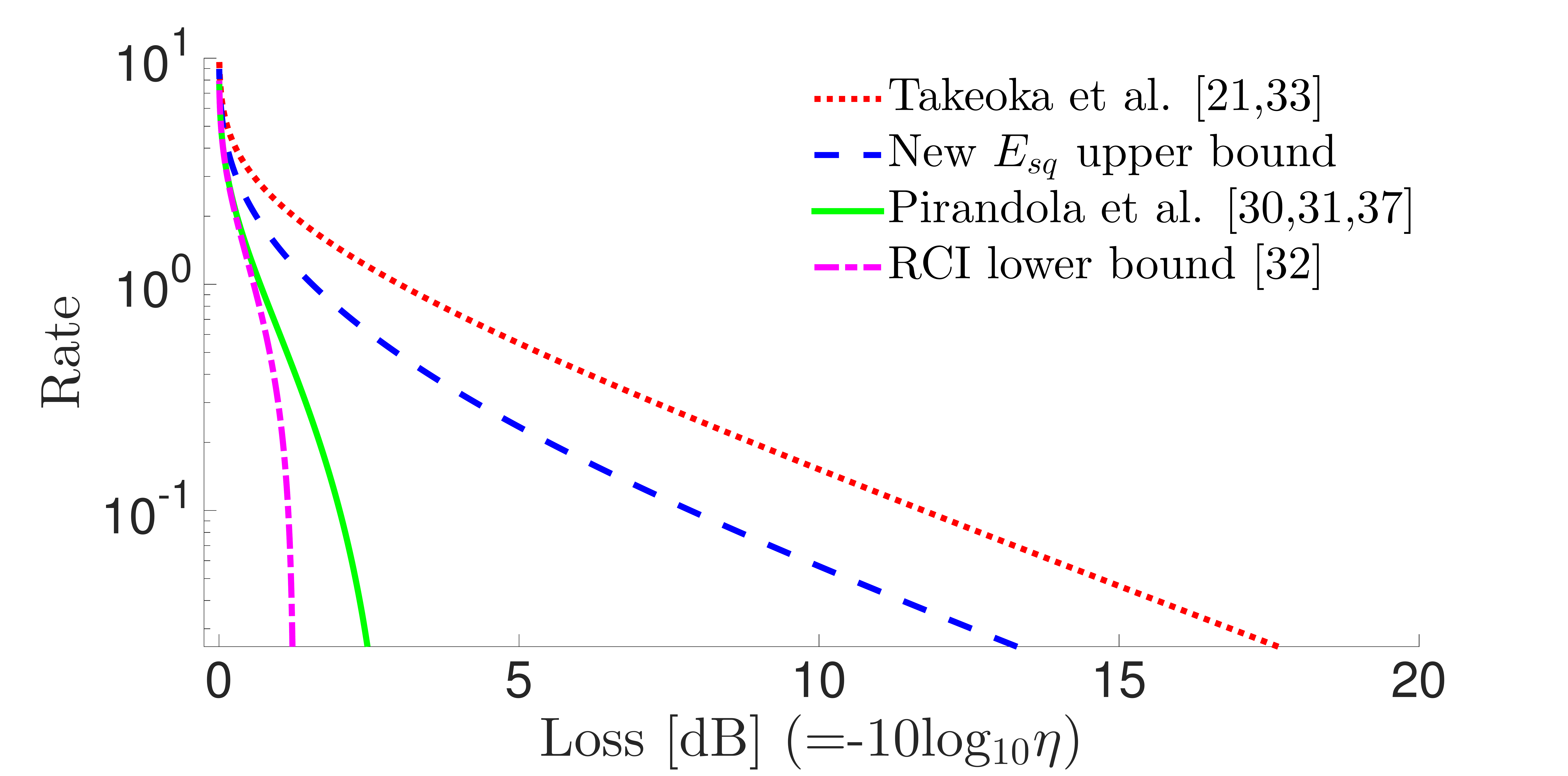}
\caption{Bounds on the squashed entanglement of the thermal channel with $N_B = 1$ as a function of the loss in dB. The red dotted line shows the upper bound by Takeoka et al.~\cite{takeoka2014squashed,takeoka2014squashed2}, the dashed blue line the new bound reported in this paper, in solid green the bound by Pirandola et al.~\cite{pirandola2015general, pirandola2015ultimate, Pirandola:2015aa}, and the dash-dotted line shows the reverse coherent information~\cite{garcia2009reverse} which is a lower bound.
}
\label{fig:Thermalplot}
\end{figure}

\noindent A thermal channel is similar to the pure-loss channel, but instead of the input interacting with a vacuum state on a beamsplitter of transmissivity $\tau$, it interacts with a thermal state with mean photon number $N_B$. For a thermal channel we have that $G = (1-\eta)N_B + 1$ and $T = \frac{\eta}{ (1-\eta)N_B + 1}$. In Figure \ref{fig:Thermalplot} the upper bound is plotted for $N_B = 1$ together with two other bounds and the reverse coherent information, which is a lower bound on $P_2(\N)$~\cite{garcia2009reverse}.

\subsubsection{Non-quantum limited noise for lossy channels}
\noindent In experimental setups one does not measure $\nu$, but the additional noise $\chi \geq 0$. We have the relation $\nu = 1-\tau + \chi$ where $1-\tau$ is the minimum amount of noise that will be introduced for a loss $\tau$ (the quantum-limited noise) ~\cite{Weedbrook:2012aa}. The upper bound from \eqref{eq:Gbound2} can then be rewritten as
\begin{gather}
\frac{\zeta(\chi+2+2\tau,\chi+2-2\tau)-\tau \zeta(\chi+4,\chi)}{(4+2\chi)(1-\tau^2)}\ .
\end{gather}

\subsection{Finite-energy bounds}
\begin{figure}[h]
	\centering
\includegraphics[clip, trim=35mm 5mm 71mm 0mm, scale = 0.10]{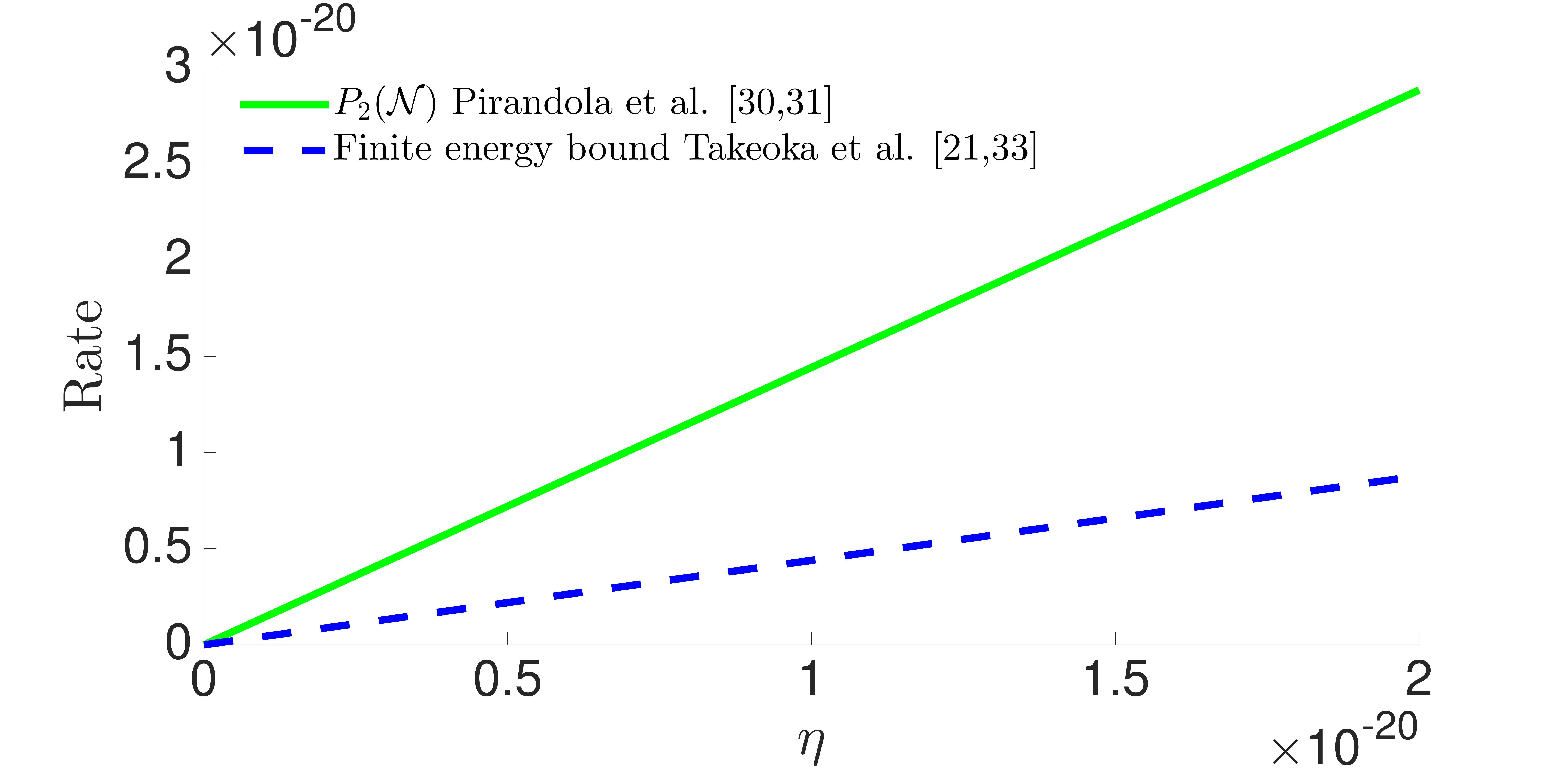}
\caption{Bound for the pure-loss channel with an average photon number of 0.1 and the secret key capacity~\cite{pirandola2015general,Pirandola:2015aa} as a function of $\eta$. The new bound in this paper coincides with the finite-energy variant of the bound by Takeoka et al., see ~\cite{takeoka2014squashed,takeoka2014fundamental}. The loss parameter $\eta$ ranges from 0 to $2\cdot 10^{-20}$, which is the range of expected losses for transmissions across fibers with length $\approx1000$ km with an attenuation length of 22 km.
}
\label{fig:finitepurelossplot}
\end{figure}

\begin{figure}[h]
	\centering
\includegraphics[clip,trim=35mm 5mm 71mm 0mm, scale = 0.10]{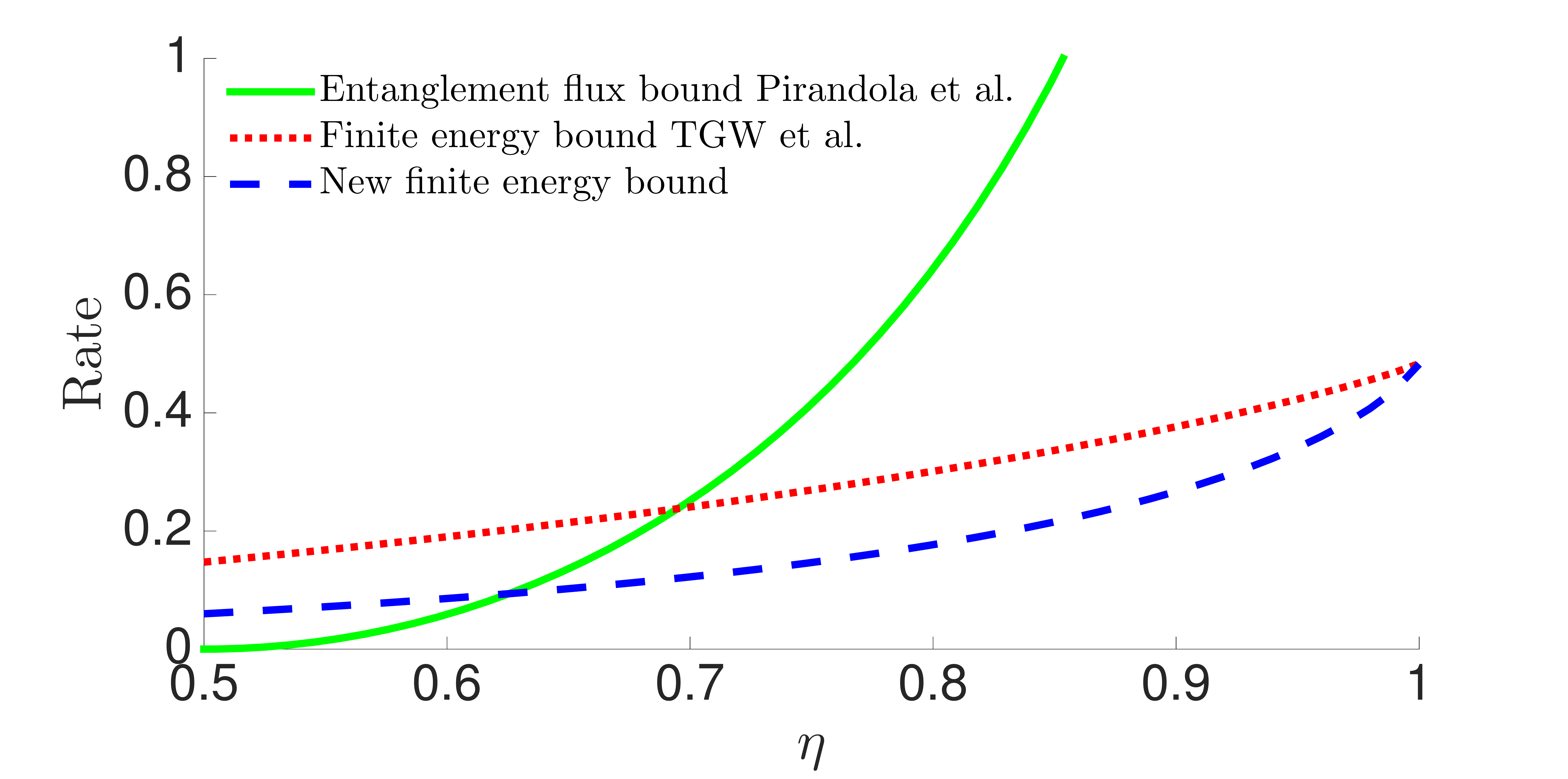}
\caption{Comparison of the upper bound found by Pirandola et al.~\cite{pirandola2015general, pirandola2015ultimate,Pirandola:2015aa} for the thermal channel with $N_B = 1$ and the two squashed entanglement finite-energy bounds with average photon number of 0.1 as a function of the loss-parameter $\eta$~\cite{takeoka2014squashed,takeoka2014squashed2}.
}
\label{fig:finitethermalplot}
\end{figure}

\noindent For low mean photon number and certain parameter ranges the finite-energy bound in equation \eqref{eq:finiten2} is tighter than previous upper bounds on the two-way assisted capacities. For any energy the pure-loss bound from Takeoka et al.~\cite{takeoka2014squashed,takeoka2014squashed2} and equation  \eqref{eq:finiten} coincide. In Figure \ref{fig:finitepurelossplot} the bound from Takeoka et al.~\cite{takeoka2014squashed,takeoka2014squashed2}, is shown for an average photon number of $N = 0.1$~\cite{benatti2010quantum,da2015linear} and the two-way assisted private capacity of the pure-loss channel~\cite{pirandola2015general, pirandola2015ultimate,Pirandola:2015aa}. The loss-parameter runs from $0$ to $2\cdot 10^{-20}$, which is the expected range of losses for fiber lengths of around 1000 kilometers.
In Figure \ref{fig:finitethermalplot} we plot the upper bound by Pirandola et al.~\cite{pirandola2015general, pirandola2015ultimate, Pirandola:2015aa}, the finite-energy bounds of Takeoka et al.~\cite{takeoka2014squashed,takeoka2014squashed2}, and equation \eqref{eq:finiten} for the thermal channel with $N_B = 1$. Note that the finite-energy bounds are zero only for $\eta = 0$, while the upper bound by Pirandola et al.~\cite{pirandola2015general, pirandola2015ultimate,Pirandola:2015aa} equals zero for $\eta \leq \frac{N_b}{N_b+1}$.

\section{Conclusion}
\noindent In this paper we have obtained bounds on the two-way assisted capacities of several relevant channels using the squashed entanglement of a quantum channel. For practical purposes, the most relevant of the channels considered are phase-insensitive Gaussian channels. Our bound for these channels is always nonzero, even when the corresponding channel is entanglement-breaking. This points to the existence of an even better squashing channel for phase-insensitive Gaussian channels. Future work could investigate this intriguing avenue, especially due to its relevance to the squashed entanglement of a bipartite state as an entanglement measure.

Furthermore, we have proven the exact two-way assisted capacities and the squashed entanglement of the $d$-dimensional erasure channel, improved the previous best known upper bound on the amplitude-damping channel and derived a squashed entanglement bound for general qubit Pauli channels. In particular, our bound applies to the depolarizing channel and improves on the previous best known squashed entanglement upper bound.

The only credible way to claim whether an implementation of a quantum repeater is good enough is by achieving a rate not possible by direct communication. Our bounds take special relevance in this context, especially for realistic energy constraints.

\section{Acknowledgements}
\noindent KG, DE and SW acknowledge support from STW, Netherlands, an ERC Starting Grant and an NWO VIDI Grant.
We would like to thank Mark M. Wilde and Stefano Pirandola for discussions regarding this project.
We also thank Marius van Eck, Jonas Helsen, Corsin Pfister, Andreas Reiserer and Eddie Schoute for helpful comments regarding an earlier version of this paper.

	      \bibliographystyle{IEEEtran}
       \bibliography{library}{}
      \onecolumngrid
       \appendix

\subsection{Bounds for convex decomposition of channels}
\label{sec:convdec}
\noindent One way of obtaining bounds on the squashed entanglement is based on decomposing the channel action as a mixture of other channels actions and bounding each of them individually. 

Let $\N_{A'\rightarrow B}$ be a channel such that its action can be written as the convex combination of the action of two other channels $\N_0$ and $\N_1$ 
\begin{equation}
\rho_{AB}=(\I\otimes\N)(\phi_{AA'})=p (\I\otimes\N_0)(\phi_{AA'}) + (1-p)(\I\otimes\N_1)(\phi_{AA'})\ .
\end{equation}
Then we can always purify $\rho_{AB}$ in the following way
\begin{equation}
\ket{\rho}_{ABEF_1F_2}=\sqrt{p}\ket{\rho^{(0)}}_{ABE}\ket{0}_{F_1}\ket{0}_{F_2}+\sqrt{1-p}\ket{\rho^{(1)}}_{ABE}\ket{1}_{F_1}\ket{1}_{F_2}
\end{equation}
where 
\begin{equation}
\ket{\rho^{(0)}}_{ABE}=V_{A'\rightarrow BE}^{\N_0}\ket{\phi}_{AA'}
\end{equation}
and 
\begin{equation}
\ket{\rho^{(1)}}_{ABE}=V_{A'\rightarrow BE}^{\N_1}\ket{\phi}_{AA'}\ .
\end{equation}
That is, $\ket{\rho^{(0)}}_{ABE}$ and $\ket{\rho^{(1)}}_{ABE}$ stand for the state that we obtain after applying the channel isometry to the pure input state $\ket{\phi}_{AA'}$. 

Let us apply the following channel to $\ket{\rho}_{ABEF_1F_2}$
\begin{equation}
\rho_{ABEF_1F_2}\mapsto \tr_{F_2}\left((\I_{AB}\otimes S^0_{E\rightarrow E'}\otimes P^{\ket{0}}_{F_1}\otimes\I_{F_2})(\rho_{ABEF_1F_2})+(\I_{AB}\otimes S^1_{E\rightarrow E'}\otimes P^{\ket{1}}_{F_1}\otimes\I_{F_2})(\rho_{ABEF_1F_2})\right)\ .
\label{eq:chform}
\end{equation} 
Where we denote by $P^{\ket{v}}_{F_1}$ the projector onto the vector $\ket{v}$. 
First we trace out $F_2$, then
\begin{equation}
{\rho}_{ABEF_1}=p\rho^{(0)}_{ABE}\otimes\ket{0}\bra{0}_{F_1}+(1-p)\rho^{(1)}_{ABE}\otimes\ket{1}\bra{1}_{F_1}\ .
\end{equation}
Now, let us apply the rest of the channel. We obtain 
\begin{equation}
\label{eq:conchstate}
{\rho}_{ABE'F_1}=\sum_iS^i_{E\rightarrow E'}\otimes\ket{i}\bra{i}_{F_1}({\rho}_{ABEF_1})=pS^0_{E\rightarrow E'}(\rho^{(0)}_{ABE})\otimes\ket{0}\bra{0}_{F_1}+(1-p)S^1_{E\rightarrow E'}(\rho^{(1)}_{ABE})\otimes\ket{1}\bra{1}_{F_1}\ .
\end{equation}
That is, ${\rho}_{ABE'F_1}$ is a quantum-classical system.  For states of this form the conditional mutual information can be simplified to
\begin{equation}\label{eq:convmutinf}
I(A;B|EF_1)= p I(A;B|E')_{S^0_{E\rightarrow E'}(\rho^{(0)}_{ABE})} + (1-p)I(A;B|E')_{S^1_{E\rightarrow E'}(\rho^{(1)}_{ABE})}
\end{equation}

\noindent Now we can upper bound $\Esq(\N)$ in the following way
\begin{align}
\Esq(\N) &\leq\max_{\phi_{AA'}}\inf_{\sum_iS^i_{E\rightarrow E'}\otimes\ket{i}\bra{i}_{F_1}\otimes\tr_{F_2}}I(A;B|E'F_1))_{{\rho}_{ABEF_1}}\\
               &=\max_{\phi_{AA'}}\left(p\inf_{S^0_{E\rightarrow E'}}I(A;B|E')_{\ket{\rho^{(0)}}_{ABE}}+(1-p)\inf_{S^1_{E\rightarrow E'}}I(A;B|E')_{\ket{\rho^{(1)}}_{ABE}}\right)\label{eq:convMutInf}\\
               &\leq p \Esq(\N_1)+(1-p)\Esq(\N_2)\ .
\end{align}
The first inequality holds by restricting the squashing channels to those channels 
of the form in \eqref{eq:chform}. Equality \eqref{eq:convMutInf} follows since for channels of the form \eqref{eq:chform} the resulting state is a quantum-classical state as indicated in \eqref{eq:conchstate}, and for classical quantum states the conditional mutual information of the whole state is a convex combination of the individual conditional mutual informations as shown in \eqref{eq:convmutinf}. The last inequality follows because the state that achieves the maximum squashed entanglement might be different for each channel. This method generalizes easily to any number of channels, from which it follows that if $\mathcal{N}(\rho) = \sum_{i}p_i\mathcal{N}_i(\rho)$ with $\sum_{i}p_i = 1$ and $p_i\geq 0$, then

\begin{gather}
Q_2(\N) \leq P_2(\N)\leq \Esq(\mathcal{N})\leq \sum_{i}p_i \Esq(\mathcal{N}_i)\ .
\end{gather}

\subsection{Improved bound for the depolarizing channel}
\label{sec:depol}
It is well known that the depolarizing channel becomes entanglement breaking for $p\geq \frac{2}{3}$~\cite{horodecki1996separability}, which implies that $P_2$ is zero in that range. For $\epsilon\leq p\leq\frac{2}{3}$, we can write the output of the channel as a convex combination of the output of $\mathcal D_{2/3}$ and $\mathcal D_{\epsilon}$. That is, there exists some $0\leq\alpha\leq 1$ such that
\begin{equation}
\label{eq:depconv}
\Dp(\rho) = (1-\alpha)\mathcal D_{\epsilon}(\rho)+\alpha\mathcal D_{2/3}(\rho).
\end{equation}
By expanding both sides of \eqref{eq:depconv} and identifying the coefficients, we obtain
\begin{equation}
\alpha=\frac{p-\epsilon}{2/3-\epsilon}\,
\end{equation}
which is in the range $[0,1]$ for $0\leq\epsilon\leq p$.

Using the decomposition of the depolarizing from \eqref{eq:depconv} the action of $\Dp$ on half of a pure entangled state takes the following form,
\begin{align}
\psi_{AB}&=\I\otimes\Dp(\ket{\psi}\bra{\psi}_{AA'})\\
              &=(1-\alpha)\left[(1-\epsilon)\ket{\psi}\bra{\psi}_{AB}+\epsilon\cdot\mms\right]+\alpha\sum_i\lambda_i\ket{\alpha_i}\bra{\alpha_i}_{A}\otimes\ket{\beta_i}\bra{\beta_i}_{B}.
\end{align}

Let $\rho_{AB}=\left((1-\epsilon)\ket{\psi}\bra{\psi}_{AB}+\epsilon\cdot\mms\right)$. A possible extension of $\psi_{AB}$ is
\begin{gather}
\psi_{ABE'} = (1-\alpha)\rho_{AB} \otimes \Ket{n+1}\Bra{n+1}_{E'} + \alpha \sum_{i=1}^{n}\lambda_i\Ket{\psi_i}\Bra{\psi_i}_A\otimes \Ket{\phi_i}\Bra{\phi_i}_B\otimes \Ket{i}\Bra{i}_{E'}.
\end{gather}
Since $\psi_{ABE'} $ is a valid extension of $\rho_{AB}$, this means that there exists some squashing channel $\mathcal{S}_{E\rightarrow E'}$ that takes the environment of the depolarizing channel to this particular $E'$. This is easy to see, first we can find a state $\ket{\psi}_{ABE'T}$ that purifies $\psi_{ABE'}$. Next, since all purifications are related by an isometry there exists some purification $V_{E\rightarrow E'T}$ that takes the environment of the channel to $E'T$. After this we trace out the system $T$ and obtain $\psi_{ABE'}$. 

Now, $\psi_{ABE'}$ is a quantum-classical system. Hence, we can decompose the conditional mutual information $I(A;B|E')$ into the sum of the mutual information conditioned on each value of $E$
\begin{align}
I(A:B|E')_{\psi} &= (1-\alpha) I(A:B|E')_{\rho}+\alpha\sum_{i=1}^n\lambda_iI(A:B|E')_{\Ket{\psi_i}\Bra{\psi_i}_A\otimes \Ket{\phi_i}\Bra{\phi_i}_B\otimes \Ket{i}\Bra{i}_{E'}}\\
			&=(1-\alpha) I(A:B|E')_{\rho}\label{eq:oneminusalpha}
\end{align}
Furthermore the input state that maximizes \eqref{eq:oneminusalpha} is the maximally entangled state on $AA'$. Hence, the following bound upper bound on $\Esq(\Dp)$ holds for $0\leq\epsilon\leq p$
\begin{equation}
\label{eq:dpinterp}
\Esq(\Dp)\leq (1-\alpha)\frac{3 \epsilon \log(\epsilon)+(4-3\epsilon)\log(4-3\epsilon)}{8}.
\end{equation}

\subsection{Squashed entanglement upper bound for any phase-insensitive Gaussian channel}
\noindent In this section we discuss a proof of an upper bound for the squashed entanglement of any phase-insensitive bosonic Gaussian channel $\mathcal{N}_{\mathrm{PI}}$. Here we use the fact that any such channel can be decomposed as a beamsplitter with transmissivity $T$ concatenated with a two-mode squeezer with squeezing parameter $r = \acosh(\sqrt{G})$. We first show that we can restrict the input states to the class of thermal states with mean photon number $N$, after which the entropic quantity of interest is written as a function of $N$. We then show that this function is monotonically increasing, after which we take the asymptotic limit $N\rightarrow \infty$ of the entropic quantity yielding

\begin{gather}
\Esq(\mathcal{N}_{\mathrm{PI}}) \leq \frac{\left(1-T^2\right)G\log(\frac{1+T}{1-T})-\left(G^2-1\right)T\log(\frac{G+1}{G-1})}{1-G^2T^2}\ .
\end{gather}
To show this is true, we first use a different form of $\Esq(\mathcal{N})$, which was proven by Takeoka et al.~\cite{takeoka2014squashed},
\begin{align}
\Esq(\mathcal{N}_{\mathrm{PI}}) & = \frac{1}{2}\max\limits_{\rho_{A'}}\inf\limits_{{V_{E\rightarrow E'F}}}\lbrack H(B|E')_{\omega}+H(B|F)_{\omega}\rbrack\ .\label{eq:sqentchannelalt}
\end{align}
There are two differences between the characterization in \eqref{eq:sqentchannelalt} and the one in \eqref{eq:squashedentchannel}. First, the maximization runs over density operators on $A'$ instead of running over pure states on $AA'$. Second, instead of taking the infimum over the squashing maps, it is taken over their dilations: squashing isometries $V_{E\rightarrow E'F}$ that take the system $E$ to $E'$ and an auxiliary system $F$. The entropies are then taken on the state $\omega$ on systems $BE'F$.

The total operation, which we denote by $\mathbf{D}$, consists of the Stinespring dilation of the channel ($\mathbf{B}_1$ and $\mathbf{S}$) and the squashing isometry consisting of two balanced beamsplitters ($\mathbf{B}_2$ and $\mathbf{B}_3$), see Figure \ref{fig:FigureAppendix}. We now write $H(B|E')=H(B|E_1E_2)$, where the system on $E'$ is the output at $E_1'$ and $E_2'$ after the total transformation $\mathbf{D}$. $E_1'$ is the state after the vacuum state on $E_1$ has interacted with the beamsplitter $\mathbf{B}_1$ and the balanced beamsplitter $\mathbf{B}_2$. Similar statements hold also for $E_2'$, $F_1'$ and $F_2'$. Since the isometry consists of two balanced beamsplitters we have that $H(B|E') = H(B|F_1'F_2') = H(B|F)$, so that $\Esq(\mathcal{N}) \leq H(B|E')$. After having found the state after the transformation we calculate the so-called symplectic eigenvalues of the states on $BE_1'E_2'$ and $E_1'E_2'$, from which we can find $H(B|E_1'E_2')$. To get an expression of the upper bound for $N\rightarrow \infty$, we calculate for three different regimes of $G$ and $T$ the asymptotic behavior of the symplectic eigenvalues, after which we show that all three regimes give rise to the same form of the upper bound.

\subsubsection{Bound for finite $N$}
\noindent A Mathematica file is included in the supplementary material to guide the reader through the calculations performed in this section. For the proof we first need to be able to calculate the entropy of a Gaussian state as a function of its covariance matrix. The entropy of an $M-$mode Gaussian state $\rho$ can be calculated by finding the $M$ symplectic eigenvalues $\nu_k\geq 1$ of the covariance matrix $\mathbf{\Gamma}$ of $\rho$ \cite{furusawa2011quantum}. It turns out that the $2M$ eigenvalues of the matrix $\mathbf{\Omega}\mathbf{\Gamma}$ are of the form $\pm i\nu_k$ \cite{de2006symplectic}, where  

\begin{gather}
\mathbf{\Omega} := \bigoplus_{k=1}^M\begin{bmatrix}
0 & 1\\
 -1 & 0
\end{bmatrix}\ .
\end{gather}
The entropy of the state is then $\sum_{k=1}^{M}g(\nu_k)$, where $g(x) = \left(\frac{x+1}{2}\right)\log(\frac{x+1}{2})-\left(\frac{x-1}{2}\right)\log(\frac{x-1}{2})$ \cite{Weedbrook:2012aa}.

\begin{figure}
\begin{centering}
\includegraphics{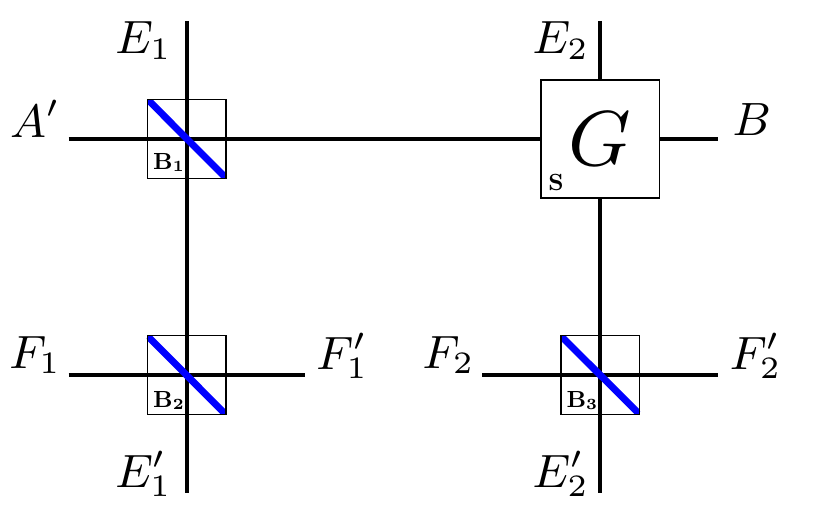}
\caption{A squashing isometry for any phase-insensitive Gaussian channel $\mathcal{N}_{\mathrm{PI}}$ taking $A'$ to $B$. The beamsplitter $\mathbf{B_1}$ and the two-mode squeezer $\mathbf{S}$ form the Stinespring dilation, while the balanced beamsplitters $\mathbf{B_2}$ and $\mathbf{B_3}$ form the squashing map. The beamsplitter $\mathbf{B}_1$ interacts with the vacuum on $E_1$ and $A$, and the two-mode squeezer $\mathbf{S}$ interacts with the output of $\mathbf{B}_2$ and the vacuum on $E_2$. The squashing isometry consists of two balanced beamsplitters $\mathbf{B_2}$ and $\mathbf{B_3}$ interacting with the vacuum on $F_1$ and $F_2$ and the output of the beamsplitter $\mathbf{B}_1$ and the two-mode squeezer $\mathbf{S}$.}
\label{fig:FigureAppendix}
\end{centering}
\end{figure}

To obtain the state at the end of the isometry we determine first the optimal state for a specified mean photon number $N$, after which we apply the Gaussian transformations of the Stinespring dilation of the channel and the isometry, shown in Figure \ref{fig:FigureAppendix}. To find the maximizing input state on $A'$, we follow the same approach~\cite{takeoka2014squashed,takeoka2014squashed2} as Takeoka et al. Since the concatenation of multiple Gaussian transformations is still a Gaussian transformation, having a Gaussian state as input will always give a Gaussian state on any of the outputs. From the extremality of Gaussian states for conditional entropy \cite{eisert2005gaussian}, we get that the optimal input state is a Gaussian state.
\newpage
\noindent To find the optimal Gaussian state, we note that the covariance matrix of all single-mode Gaussian states can be written as \cite{quantumcomm}

\begin{align}
\left(1+2N\right)\begin{bmatrix}
\cosh 2r + \cos \theta \sinh 2 r & \sin \theta \sinh 2r\\
\sin \theta \sinh 2r & \cosh 2r - \cos \theta \sinh 2 r 
\end{bmatrix}
\label{eq:generalgaussian}
\end{align}
for some $r\geq 0$ and $\theta \in \mathbb{R}$. Since the channel from $A'$ to $BE_1'E_2'F_1'F_2'$ is covariant with displacements and all unitaries $\tilde{U}$ such that the corresponding symplectic matrices $S_{\tilde{U}}$ act on the thermal state as 

\begin{align}
S_{\tilde{U}} \left(1+2N\right)\mathds{I}~{S_{\tilde{U}}}^T \rightarrow \left(1+2N\right)\begin{bmatrix}
\cosh 2r + \cos \theta \sinh 2 r & \sin \theta \sinh 2r\\
\sin \theta \sinh 2r & \cosh 2r - \cos \theta \sinh 2 r 
\end{bmatrix} \ ,
\end{align}
we have that $H(B|E')_{\rho} = H(B|E')_{\tilde{U}\rho \tilde{U}^{\dagger}}$. We set $\rho$ equivalent to $\tilde{U}\rho\tilde{U}^{\dagger}$, defining an equivalence relation. It is clear that all states with fixed  $N$ in equation \eqref{eq:generalgaussian} define an equivalence class with respect to the equivalence relation. Since $H(B|E')_{\rho} = H(B|E')_{\tilde{U}\rho \tilde{U}\dagger}$, we can set the thermal state $\left(1+2N\right)\mathds{I}$ to be the representative of that equivalence class, and we only have to consider thermal states for the optimization.\\

\noindent The total system $\Gamma_{A'E'_1F_1E'_2F_2}$ consists then of a thermal state $\Gamma_{A'}$ with mean photon number $N$ on $A'$ and vacuum states on all the other inputs:
        
\begin{gather}
\mathbf{\Gamma}_{A'E_1F_1E_2F_2} = \mathbf{\gamma}_{A'} \oplus \mathds{I}_{E_1}\oplus \mathds{I}_{F_1}\oplus \mathds{I}_{E_2}\oplus \mathds{I}_{F_2},\\
\gamma_{A'} = \begin{bmatrix}
1+2N & 0\\
0 & 1+2N
\end{bmatrix}.
\end{gather}

\noindent The operations of the isometry are then the first beamsplitter $\mathbf{B_1}$ with transmissivity $T$ on $A'$ and $E_1$

\begin{gather}
\mathbf{B_1} = 
{\begin{bmatrix}
\sqrt{T} & 0 & \sqrt{1-T} & 0\\
0 & \sqrt{T} & 0 & \sqrt{1-T}\\
-\sqrt{1-T} & 0 & \sqrt{T} & 0\\
0 & -\sqrt{1-T} & 0 & \sqrt{T}\\
\end{bmatrix}}_{A'E'_1}\oplus \mathds{I}_{F_1}\oplus \mathds{I}_{E_2}\oplus \mathds{I}_{F_2},
\end{gather}

\noindent the second beamsplitter $\mathbf{B_2}$ with transmissivity $\frac{1}{2}$ on $E_1$ and $F_1$

\begin{gather}
\mathbf{B_2} = 
\mathds{I}_{A'}\oplus
{\begin{bmatrix}
\frac{1}{\sqrt{2}} & 0 & \frac{1}{\sqrt{2}} & 0\\
0 & \frac{1}{\sqrt{2}} & 0 & \frac{1}{\sqrt{2}}\\
-\frac{1}{\sqrt{2}} & 0 & \frac{1}{\sqrt{2}} & 0\\
0 & -\frac{1}{\sqrt{2}} & 0 & \frac{1}{\sqrt{2}}\\
\end{bmatrix}}_{E_1F_1}\oplus \mathds{I}_{E_2}\oplus \mathds{I}_{F_2},
\end{gather}

\noindent the two-mode squeezer $\mathbf{S}$ on $A'$ and $E_2$ with the relation $G = \cosh^2(r)$

\begin{gather}
\mathbf{S} = 
\begin{bmatrix}
\sqrt{G} & 0 & 0 & 0 & 0 & 0  & \sqrt{G-1} & 0\\
0 & \sqrt{G} & 0 & 0 & 0 & 0 & 0 &   -\sqrt{G-1}\\
0 & 0 & 1 & 0 & 0 & 0 & 0 & 0\\
0 & 0 & 0 & 1 & 0 & 0 & 0 & 0\\
0 & 0 & 0 & 0 & 1 & 0 & 0 & 0\\
0 & 0 & 0 & 0 & 0 & 1 & 0 & 0\\
\sqrt{G-1} & 0 & 0 & 0 & 0 & 0  & \sqrt{G} & 0\\
0 & -\sqrt{G-1} & 0 & 0 & 0 & 0 & 0 &  \sqrt{G}\\
\end{bmatrix}_{A'E'_1F_1'}\oplus \mathds{I}_{F_2},
\end{gather}

\noindent and finally the last beamsplitter $\mathbf{B_3}$ on $E_2$ and $F_2$ with transmissivity $\frac{1}{2}$

\begin{gather}
\mathbf{B_3} =
\mathds{I}_{A'}\oplus \mathds{I}_{E'_1}\oplus \mathds{I}_{F_1}\oplus
{\begin{bmatrix}
\frac{1}{\sqrt{2}} & 0 & \frac{1}{\sqrt{2}} & 0\\
0 & \frac{1}{\sqrt{2}} & 0 & \frac{1}{\sqrt{2}}\\
-\frac{1}{\sqrt{2}} & 0 & \frac{1}{\sqrt{2}} & 0\\
0 & -\frac{1}{\sqrt{2}} & 0 & \frac{1}{\sqrt{2}}\\
\end{bmatrix}}_{E'_2F_2}.
\end{gather}

We then have that the total symplectic transformation matrix $\mathbf{D}$ is

\begin{gather}
\mathbf{D} = \mathbf{B_3}\mathbf{S}\mathbf{B_2}\mathbf{B_1}\\
= \begin{bmatrix}
\sqrt{GT}  & 0 & \sqrt{G(1-T)} & 0 & 0 & 0 & \sqrt{G-1} & 0 & 0 & 0\\ 
0 & \sqrt{GT}  & 0 & \sqrt{G(1-T)} & 0 & 0 & 0 & -\sqrt{G-1} & 0 & 0\\
-\sqrt{\frac{1-T}{2}} & 0 &  \sqrt{\frac{T}{2}} & 0 & \frac{1}{\sqrt{2}} & 0 & 0 & 0 & 0 & 0\\ 
0 & -\sqrt{\frac{1-T}{2}} & 0 &  \sqrt{\frac{T}{2}} & 0 & \frac{1}{\sqrt{2}} & 0 & 0 & 0 & 0\\
\sqrt{\frac{1-T}{2}} & 0 &  -\sqrt{\frac{T}{2}} & 0 & \frac{1}{\sqrt{2}} & 0 & 0 & 0 & 0 & 0\\ 
0 & \sqrt{\frac{1-T}{2}} & 0 &  -\sqrt{\frac{T}{2}} & 0 & \frac{1}{\sqrt{2}} & 0 & 0 & 0 & 0\\
\sqrt{\frac{(G-1)T}{2}} & 0 & \sqrt{\frac{(G-1)(1-T)}{2}} & 0 & 0 & 0 & \sqrt{\frac{G}{2}} & 0 & \frac{1}{\sqrt{2}} & 0\\
0 & -\sqrt{\frac{(G-1)T}{2}} & 0 & -\sqrt{\frac{(G-1)(1-T)}{2}} & 0 & 0 & 0 & \sqrt{\frac{G}{2}} & 0 & \frac{1}{\sqrt{2}}\\
-\sqrt{\frac{(G-1)T}{2}} & 0 & -\sqrt{\frac{(G-1)(1-T)}{2}} & 0 & 0 & 0 & -\sqrt{\frac{G}{2}} & 0 & \frac{1}{\sqrt{2}} & 0\\
0 & \sqrt{\frac{(G-1)T}{2}} & 0 & \sqrt{\frac{(G-1)(1-T)}{2}} & 0 & 0 & 0 & -\sqrt{\frac{G}{2}} & 0 & \frac{1}{\sqrt{2}}\\
\end{bmatrix}\ .
\end{gather}
\noindent The covariance matrix $\mathbf{\Gamma}_{BE_1'F_1'E_2'F_2'} = \mathbf{D}\mathbf{\Gamma}_{A'E_1F_1E_2F_2}\mathbf{D}^T$ after the transformation is then

\begin{gather}
\begin{bmatrix}
a\mathds{I} & -b\mathds{I} & b\mathds{I} & c\sigma_z & -c\sigma_z\\
-b\mathds{I} & d\mathds{I} & e \mathds{I} & -f\sigma_z & f\sigma_z\\
b\mathds{I} & e \mathds{I} & d\mathds{I} & f\sigma_z & -f\sigma_z\\
c\sigma_z & -f\sigma_z & f\sigma_z & g\mathds{I} & h\mathds{I}\\
-c\sigma_z & f \sigma_z & -f\sigma_z & h\mathds{I} & g\mathds{I}
\end{bmatrix},
\end{gather}
where $\sigma_z = \begin{bmatrix}1 & 0\\
0 & -1
\end{bmatrix}$, $\mathds{I} = \begin{bmatrix}
1 & 0\\
0 & 1
\end{bmatrix}$ and 

\begin{align}
a &= 2G(1+NT)-1\\
b &= N\sqrt{2\left(GT\left(1-T\right)\right)}\\
c &= \left(1+NT\right)\sqrt{2\left(G-1\right)G}\\
d &= 1+N\left(1-T\right)\\
e &= N\left(T-1\right)\\
f &= N\sqrt{\left(G-1\right)\left(1-T\right)T}\\
g &= G+\left(G-1\right)NT\\
h &= -(G-1)(1+NT).
\end{align}

\noindent The covariance matrix on the subsystems $E_1E_2$ is then

\begin{gather}
\mathbf{\Gamma}_{E_1'E_2'}=
\begin{bmatrix}
d\mathds{I} & -f\sigma_z\\
-f\sigma_z & g\mathds{I}
\end{bmatrix}.
\end{gather}
Multiplying by $\mathbf{\Omega}$ gives
\begin{gather}
\mathbf{\Omega}\mathbf{\Gamma}_{E_1'E_2'} =
\begin{bmatrix}
0  & d & 0 & f\\
-d & 0 & f & 0\\
0 & f & 0 & g\\
f & 0 & -g &0
\end{bmatrix}.
\end{gather}

Now set $\Omega^{\pm} = \sqrt{(1+N)^2-4NT\pm 2G(1+N)(NT-1)+(G+GNT)^2}$. Taking the covariance matrix corresponding to $E'_1E'_2$ we find using Mathematica the symplectic eigenvalues to be
\begin{gather}
\left(\nu_{E_1'E_2'}\right)_{1} = \left|\sqrt{-\frac{1+G^2+2N\left(1-T+GT\left(G-1\right)\right)+N^2\left(GT-1\right)^2+\left(G-1+N(GT-1)\right)\Omega^-}{2}}\right| \label{eq:eig1}\\
\left(\nu_{E_1'E_2'}\right)_{2} = \left|\sqrt{-\frac{1+G^2+2N\left(1-T+GT\left(G-1\right)\right)+N^2\left(GT-1\right)^2-\left(G-1+N(GT-1)\right)\Omega^-}{2}}\right|\label{eq:eig2}.
\end{gather}

\noindent The covariance matrix corresponding to $BE_1'E_2'$ is
\begin{gather}
\mathbf{\Gamma}_{BE_1'E_2'} =
\begin{bmatrix}
a\mathds{I} & -b\mathds{I} & c\sigma_z\\
-b\mathds{I} & d\mathds{I} & -f\sigma_z\\
c\sigma_z & -f\sigma_z & g\mathds{I}
\end{bmatrix},
\end{gather}
so that
\begin{gather}
\mathbf{\Omega}\mathbf{\Gamma}_{BE_1'E_2'} =
\begin{bmatrix}
0 & a & 0 & -b & 0 & -c\\
-a & 0 & b & 0 & -c & 0\\
0 & -b & 0 & d & 0 & f\\
b & 0 & -d & 0 & f & 0\\
0 & -c & 0 & f & 0 & g\\
-c & 0 & f & 0 & -g & 0
\end{bmatrix},
\end{gather}

\noindent From this the symplectic eigenvalues can be calculated to be

\begin{gather}
\left(\nu_{BE_1'E_2'}\right)_{1} = \left|\sqrt{-\frac{1+G^2+2N\left(1-T+GT\left(G+1\right)\right)+N^2\left(1+GT\right)^2+\left(1+G+N(1+GT)\right)\Omega^+}{2}}\right| \label{eq:eig3}\\
\left(\nu_{BE_1'E_2'}\right)_{2} = \left|\sqrt{-\frac{1+G^2+2N\left(1-T+GT\left(G+1\right)\right)+N^2\left(1+GT\right)^2-\left(1+G+N(1+GT)\right)\Omega^+}{2}}\right| \label{eq:eig4}\\
\left(\nu_{BE_1'E_2'}\right)_{3} = 1 \label{eq:eig5}.
\end{gather}

\noindent We can now calculate $H(B|E_1'E_2')$,
\begin{align}
H(B|E_1'E_2') &= H(BE_1'E_2')-H(E_1'E_2')\\
&= g\left(\left(\nu_{BE_1'E_2'}\right)_{1}\right) + g\left(\left(\nu_{BE_1'E_2'}\right)_{2}\right) - g\left(\left(\nu_{E_1'E_2'}\right)_{1}\right) - g\left(\left(\nu_{E_1'E_2'}\right)_{2}\right),
\label{eq:finiten}
\end{align}
where we used that $g(1) = 0$.

\newpage
\subsubsection{Monotonicity of the bound}
\noindent For this section we restrict ourselves to the picture of calculating the squashed entanglement on the systems $ABE_1'E_2'$ instead of $BE_1'E_2'F_1'F_2'$, where $V_{A'\rightarrow BE_1'E_2'F_1'F_2'} := V$ is the total isometry (see Figure \ref{fig:FigureAppendix}). In this picture the optimization is over the purification of the thermal state, the two-mode squeezed vacuum state $\Psi^N$. To show monotonicity of equation \eqref{eq:finiten} in $N$, we use that, up to a displacement on $B$ (conditioned on a measurement outcome $k$ at $A'$), it is possible to transform the state $\Psi^N_{AA'}$ to $\Psi^{N'}_{AA'}$, using a local operation $\Lambda_A$ on Alice (where $N'<N$)~\cite{giedke2003entanglement}.

Suppose now that $A$ performs the operation $\Lambda_A$ on the state $\rho^N_{ABE_1'E_2'} := \tr_{F_1'F_2'}\left(V\Psi^NV^{\dagger}\right)$ after the isometry,

\vspace{-4.5mm}
\begin{centering}

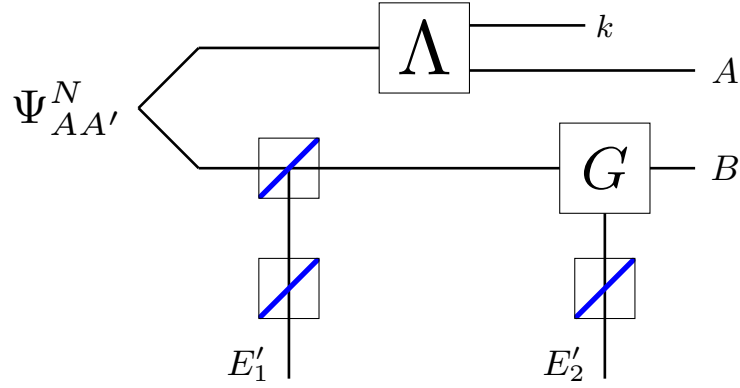
\begin{figure}
\begin{centering}
\begin{tikzpicture}[ scale=.4][H!]

	\path[draw,line width= 1 pt] (0,0) -- (-2,2); 
	\path[draw,line width= 1 pt] (-2,2) -- (0,4); 
	\path[draw,line width= 1 pt] (0,4) -- (6,4); 
	

	\draw (6,4.5-2) rectangle (9,7.5-2);
	\path[draw,line width= 1 pt] (9,6+0.75-2) -- (12.85,6+0.75-2); 
	\node[draw=none,scale=1.35] at (13.5,6.75-2.) {$k$};
	\node[draw=none,scale=1.5] at (17.5,5.25-2) {$A$};
	\path[draw,line width= 1 pt] (9,6-2.75) -- (16.5,6-2.75); 

\node[draw=none,scale=2.5] at (13.5,0) {$G$};
\node[draw=none,scale=1.5] at (1.65,-8.5+2) {$E_1'$};
\node[draw=none,scale=1.5] at (10.5+1.65,-8.5+2) {$E_2'$};
\node[draw=none,scale=2.0] at (-4.35,2) {$\Psi^{N}_{AA'}$};
\node[draw=none,scale=3.25] at (7.5,6.0-2) {$\Lambda$};
\node[draw=none,scale=1.5] at (17.5,0.0) {$B$};

	\path[draw,line width= 1 pt] (0,0) -- (12,0); 
	\path[draw,line width= 1 pt] (15,0) -- (16.5,0); 
	\path[draw,line width= 1 pt] (3,0) -- (3,-7); 
	\path[draw,line width= 2 pt, color = blue, line cap = round] (2+0.065,-1+0.065)-- (4-0.065,1-0.065); 
	\draw (2,-1) rectangle (4,1);
	\path[draw,line width= 2 pt, color = blue, line cap = round] (2+0.065,-7+0.065+2)-- (4-0.065,-5-0.065+2); 
	\draw (2,-7+2) rectangle (4,-5+2);
	\draw (12,1.5) rectangle (15,-1.5);	
	\path[draw,line width= 1 pt] (13.5,-1.5) -- (13.5,-7); 
	\path[draw,line width= 2 pt, color = blue,line cap = round] (12.5+0.065,-7+0.065+2)-- (14.5-0.065,-5-0.065+2); 
	\draw (12.5,-7+2) rectangle (14.5,-5+2);
\end{tikzpicture}
\caption{Alice can perform a local operation $\Lambda$ on one half of $\Psi^{N}_{AA'}$ that yields a state on $A$ and a classical outcome $k$. The state conditioned on the outcome $k$ on systems $ABE_1'E_2$ is, up to a unitary displacement on $B$ and $E_1'E_2'$, equal to the state $\rho^{N'}$. Alice and Bob can thus simulate any lower energy scenario.}
\end{centering}
\label{fig:figteleport}
\end{figure}
\end{centering}
\begin{align}
\left(\Lambda_A \otimes \mathds{I}_{BE_1'E_2'}\right)\rho^N_{ABE_1'E_2'} &= \int\hspace{-0.7mm}\mathrm{d}k\Ket{k}\hspace{-0.7mm}\Bra{k} \otimes \left(\left(\mathds{I}_A \otimes U^k_B\otimes U^k_{E_1'E_2'}\right)\rho^{N'}_{ABE_1'E_2'}\right)\\
& = \int\hspace{-0.7mm}\mathrm{d}k \Ket{k}\hspace{-0.7mm}\Bra{k} \otimes \rho^{N'\hspace{-0.6mm},\hspace{0.5mm}k}_{ABE_1'E_2'}\ .
\end{align}
Here we used that displacement operations can always be removed by local operations~\cite{holevo2012quantum}, so that for fixed outcome $k$ the state $\rho^{N'\hspace{-0.6mm},\hspace{0.5mm}k}_{ABE_1'E_2'}$ is related to $\rho^{N'}_{ABE_1'E_2'}:= \tr_{F_1'F_2'}\left(V\Psi^{N'}\right)$ by unitary displacements on $B$ and $E_1'E_2'$. The conditional mutual information evaluated on the state $\Lambda_A \otimes \mathds{I}_{BE_1'E_2'}\rho^N_{ABE_1'E_2'} = \tilde{\rho}^N$ then satisfies
\begin{align}
I(A;B|E_1'E_2')_{\rho^N_{ABE_1'E_2'}}&\geq I(A;B|E_1'E_2')_{\tilde{\rho}^N}\label{eq:cmi1}\\
&\geq \int\mathrm{d}k~I(A;B|E_1'E_2')_{\rho^{N'\hspace{-0.6mm},\hspace{0.5mm}k}}\label{eq:cmi2}\\
& = \int\mathrm{d}k~I(A;B|E_1'E_2')_{\rho^{N'}}\label{eq:cmi3}\\
& =I(A;B|E_1'E_2')_{\rho^{N'}}\label{eq:cmi4}\ .
\end{align}
In equation \eqref{eq:cmi1} we used that the conditional mutual information can never increase under local operations on $A$~\cite{christandl2004squashed}. In equation \eqref{eq:cmi2} we use the fact that the states $\rho^{N'\hspace{-0.6mm},\hspace{0.5mm}k}_{ABE_1'E_2'}$ are flagged on the classical outcome $k$, and that the conditional mutual information of the whole state can not be smaller than the sum of the values of the conditional mutual information of the individual states~\cite{christandl2004squashed}. In equations \eqref{eq:cmi3} and \eqref{eq:cmi4} we use the fact that all the $\rho^{N'\hspace{-0.6mm},\hspace{0.5mm}k}_{ABE_1'E_2'}$ states are related to $\rho^{N'}_{ABE_1'E_2'}$ by local unitaries on $B$ and $E_1'E_2'$ and that the conditional mutual information of those states thus must be equal.

That is, the conditional mutual information computed over the isometry $V$ with input state $\Psi^N$ is always greater than the conditional mutual information computed over the isometry $V$ with input state $\Psi^{N'}$ if $N'<N$. This thus implies that equation \eqref{eq:finiten} is a bound for all phase-insensitive Gaussian bosonic channels and all energy restrictions.

\subsubsection{Expression as $N\rightarrow \infty$}
\noindent To obtain an explicit form for the expression in \eqref{eq:finiten} as $N\rightarrow \infty$, we expand the eigenvalues around $N = \infty$ for three different regimes of $G$ and $T$ using Mathematica. For $G = \frac{1}{T}$ we have

\begin{align}
\left(\nu_{E_1'E_2'}\right)_{1} &= \sqrt{\frac{G^2-1}{G}}\sqrt{N}+\mathcal{O}\left(1\right),\\
\left(\nu_{E_1'E_2'}\right)_{2}&= \sqrt{\frac{G^2-1}{G}}\sqrt{N}+\mathcal{O}\left(1\right),\\
\left(\nu_{BE_1'E_2'}\right)_{1} &= 2N+\mathcal{O}\left(1\right),\\
\left(\nu_{BE_1'E_2'}\right)_{2} &= \frac{G^2+1}{2G}+o\left(1\right)\ .
\end{align}
Here we used the notation that $f(N) = o(h(N))$ for two functions $f(N)$ and $h(N)$ if and only if $\forall \epsilon>0, \exists N'$ such that $\forall N>N',~f(N)\leq \epsilon h(N)$.\\

\noindent Now let us introduce the equivalence relation $\rightleftarrows$ for two functions $f(N)$ and $h(N)$, so that $f(N)\rightleftarrows h(N)$ if and only if $\lim_{N\rightarrow \infty}\left|f(N)-h(N)\right| = 0$, i.e. we can safely replace $f(N)$ by $h(N)$ as $N\rightarrow \infty$. For example, we have that $g(N+c) \rightleftarrows g(N) \rightleftarrows \log(\frac{N}{2})+\frac{1}{\ln2}$. In particular, if $f(N) = h(N) + o(1)$, then $g(f(N)) \rightleftarrows g(h(N))$. Furthermore, this also means that if we have $f(N) = h(N)+\mathcal{O}(1)$ and $\lim_{N\rightarrow \infty}f(N) = \lim_{N\rightarrow \infty} h(N) = \infty$, then $g(f(N)) \rightleftarrows \log(\frac{h(N)}{2}) + \frac{1}{\ln2}$. We will call these relations the asymptotic entropic relations for short.\\

\noindent Using these asymptotic entropic relations, we find
\begin{gather}
H(BE_1'E_2')-H(E_1'E_2') = g\left(\left(\nu_{BE_1E_2}\right)_{1}\right)+g\left(\left(\nu_{BE_1E_2}\right)_{2}\right)-g\left(\left(\nu_{E_1E_2}\right)_{1}\right)-g\left(\left(\nu_{E_1E_2}\right)_{2}\right) \label{eq:step1} \\
\rightleftarrows g\left(\frac{G^2+1}{2G}\right)+g\left(2N\right)-g\left( \sqrt{\frac{G^2-1}{G}}\sqrt{N}\right)-g\left( \sqrt{\frac{G^2-1}{G}}\sqrt{N}\right) \label{eq:step2} \\
\rightleftarrows g\left(\frac{G^2+1}{2G}\right) + \log(N)+\frac{1}{\ln2}-\log(\sqrt{\frac{G^2-1}{4G}}\sqrt{N})-\frac{1}{\ln2}-\log(\sqrt{\frac{G^2-1}{4G}}\sqrt{N})-\frac{1}{\ln2} \label{eq:step3}\\
=  g\left(\frac{G^2+1}{2G}\right)+\log(\frac{4G}{G^2-1})-\frac{1}{\ln2} \label{eq:step4}\\
= \frac{\frac{G^2+1}{2G}+1}{2}\log(\frac{\frac{G^2+1}{2G}+1}{2})-\frac{\frac{G^2+1}{2G}-1}{2}\log(\frac{\frac{G^2-1}{2G}+1}{2})+\log(\frac{4G}{G^2-1})-\frac{1}{\ln2} \label{eq:step5}\\
= \frac{\left(G+1\right)^2}{4G}\log(\frac{\left(G+1\right)^2}{4G})-\frac{\left(G-1\right)^2}{4G}\log(\frac{\left(G-1\right)^2}{4G})+\log(\frac{4G}{G^2-1})-\frac{1}{\ln2} \label{eq:step6} \\
=  \frac{\left(G+1\right)^2}{4G}\log(\left(G+1\right)^2)-\frac{\left(G-1\right)^2}{4G}\log(\left(G-1\right)^2) \nonumber \\
+\underbrace{\left(-\frac{\left(G+1\right)^2}{4G}+\frac{\left(G-1\right)^2}{4G}+1\right)}_{0}\log(4G)-\log(G^2-1)-\frac{1}{\ln2} \label{eq:step7} \\
=  \frac{\left(G+1\right)^2}{2G}\log(G+1)-\frac{\left(G-1\right)^2}{2G}\log(G-1)-\log(G^2-1)-\frac{1}{\ln2} \label{eq:step8} \\
= \left(\frac{G^2+2G+1}{2G}\right)\log(G+1)-\left(\frac{G^2-2G+1}{2G}\right)\log(G-1)-\log(G^2-1)-\frac{1}{\ln2} \label{eq:step9}\\
= \frac{G^2+1}{2G}\log(\frac{G+1}{G-1}) - \underbrace{\log(G+1)+\log(G-1)-\log(G^2-1)}_{0}-\frac{1}{\ln2} \label{eq:step10}\\
= \frac{G^2+1}{2G}\log(\frac{G+1}{G-1})-\frac{1}{\ln2} = \frac{T^2+1}{2T}\log(\frac{1+T}{1-T})-\frac{1}{\ln2}.
\label{eq:BoundForGEqualT}
\end{gather}

\noindent Here we used the asymptotic entropic relations in equations \eqref{eq:step2} and \eqref{eq:step3}. Equation \eqref{eq:step4} is basic rewriting, equation \eqref{eq:step5} follows directly from the definition of $g(\cdot)$, and equation \eqref{eq:step6} follows from rewriting the terms. In equation \eqref{eq:step7} we collect the terms proportional to $\log(4G)$, from which we can see that these terms sum up to zero. In equation \eqref{eq:step9} we expand the quadratic terms, collect corresponding terms in equation \eqref{eq:step10} and write the upper bound both as a function of $G$ and $T$ in the last equality.\\

\noindent For $G > \frac{1}{T}$ we get in the asymptotic limit that equations \eqref{eq:eig1}, \eqref{eq:eig2}, \eqref{eq:eig3} and \eqref{eq:eig4} become

\begin{align}
\left(\nu_{E_1'E_2'}\right)_{1} &= N\left(GT-1\right)+\mathcal{O}\left(1\right),\\
\left(\nu_{E_1'E_2'}\right)_{2} &= \frac{G-T}{GT-1}+o\left(1\right),\\
\left(\nu_{BE_1'E_2'}\right)_{1} &= N\left(1+GT\right)+\mathcal{O}\left(1\right),\\
\left(\nu_{BE_1'E_2'}\right)_{2} &= \frac{G+T}{1+GT}+o\left(1\right).
\end{align}

\noindent For $G<\frac{1}{T}$ we have

\begin{align}
\left(\nu_{E_1'E_2'}\right)_{1} &= \frac{G-T}{1-GT}+o\left(1\right),\\
\left(\nu_{E_1'E_2'}\right)_{2} &= N\left(1-GT\right)+\mathcal{O}\left(1\right),\\
\left(\nu_{BE_1'E_2'}\right)_{1} &= N\left(1+GT\right)+\mathcal{O}\left(1\right),\\
\left(\nu_{BE_1'E_2'}\right)_{2} &= \frac{G+T}{1+GT}+o\left(1\right)\ .
\end{align}

\noindent For both regimes, the eigenvalues and in particular their leading terms are always positive. We see that for both $G>\frac{1}{T}$ and $G<\frac{1}{T}$ the absolute value of the eigenvalues are the same up to ordering, so that

\begin{gather}
H(BE_1'E_2')-H(E_1'E_2') \rightleftarrows g\left(\frac{G+T}{1+GT}\right)+g\left(N\left(1+GT\right)\right)-g\left(N\left|1-GT|\right)\right)-g\left(\frac{G-T}{\left|1-GT\right|}\right)\\
\rightleftarrows g\left(\frac{G+T}{1+GT}\right) + \log(\frac{N\left(1+GT\right)}{2})+\frac{1}{\ln2}-\log(\frac{N\left|1-GT\right|}{2})-\frac{1}{\ln2} - g\left(\frac{G-T}{\left|1-GT\right|}\right)\\
= g\left(\frac{G+T}{1+GT}\right)-g\left(\frac{G-T}{\left|1-GT\right|}\right)+\log(\frac{1+GT}{\left|1-GT\right|})
\label{formula2step1}\\
= g\left(\frac{G+T}{1+GT}\right)-g\left(\frac{G-T}{1-GT}\right)+\log(\frac{1+GT}{1-GT})\ ,
\label{eq:sqeunfinished}
\end{gather}

\noindent where in the first and second step we again used the asymptotic entropic relations. Equation \eqref{formula2step1} is basic algebraic rewriting of the logarithms. We can drop the absolute signs going from equation \eqref{formula2step1} to \eqref{eq:sqeunfinished}. To see this, note that $\log(-x) = \log(x)+\frac{i\pi}{\ln2}$ for $x>0$, where we choose the branch cut along the negative imaginary axis, and in a similar way we find that $g(-y) = \frac{-y+1}{2}\log(\frac{-y+1}{2})-\frac{-y-1}{2}\log(\frac{-y-1}{2}) = \frac{y+1}{2}\log(-\frac{y+1}{2})-\frac{y-1}{2}\log(-\frac{y-1}{2}) =  g(y)+\frac{i\pi}{\ln2}$ for $y\geq 1$. From this we find that $-g(-y)+\log(-x) = -g(y)+\log(x)$ for $x>0,~y \geq 1$. Since $\frac{G-T}{\left|1-GT\right|}> 1$ and $\frac{1+GT}{\left|1-GT\right|}\geq 0$ for $G\geq 1,~0\leq T \leq 1$, we have that $-g\left(\frac{G-T}{\left|1-GT\right|}\right)+\log(\frac{1+GT}{\left|1-GT\right|}) = -g\left(\frac{G-T}{1-GT}\right)+\log(\frac{1+GT}{1-GT})$.
\newpage
\noindent We can rewrite equation \eqref{eq:sqeunfinished} as

\begin{gather}
g\left(\frac{G+T}{1+GT}\right)-g\left(\frac{G-T}{1-GT}\right)+\log(\frac{1+GT}{1-GT})\\
= \frac{\frac{G+T}{1+GT}+1}{2}\log(\frac{\frac{G+T}{1+GT}+1}{2})-\frac{\frac{G+T}{1+GT}-1}{2}\log(\frac{\frac{G+T}{1+GT}-1}{2})-\frac{\frac{G-T}{1-GT}+1}{2}\log(\frac{\frac{G-T}{1-GT}+1}{2}) \nonumber \\
-\frac{\frac{G-T}{1-GT}-1}{2}\log(\frac{\frac{G-T}{1-GT}-1}{2})+\log(\frac{1+GT}{1-GT})\\
=  \frac{(G+1)(1+T)}{2\left(1+GT\right)}\log(\frac{(G+1)(1+T)}{2\left(1+GT\right)})-\frac{(G-1)(1-T)}{2\left(1+GT\right)}\log(\frac{(G-1)(1-T)}{2\left(1+GT\right)}) \nonumber \\
- \frac{(G+1)(1-T)}{2\left(1-GT\right)}\log(\frac{(G+1)(1-T)}{2\left(1-GT\right)})+\frac{(G-1)(1+T)}{2\left(1-GT\right)}\log(\frac{(G-1)(1+T)}{2\left(1-GT\right)})+\log(\frac{1+GT}{1-GT})\ ,
\label{eq:sqeunfinished2}
\end{gather}
where we have used the definition of $g(\cdot)$ in the first equality and simplified the terms in the second step.

We can expand the logarithms and collect the different terms and simplify to rewrite equation \eqref{eq:sqeunfinished2}. 
Let us consider one by one the terms proportional to each logarithmic term. The terms proportional to $\log(G+1)$ are
\begin{gather}
\frac{\left(G+1\right)\left(1+T\right)}{2\left(1+GT\right)}-\frac{\left(G+1\right)\left(1-T\right)}{2\left(1-GT\right)}\\
= -\frac{\left(G^2-1\right)T}{1-G^2T^2}\ ,
\end{gather}
the terms proportional to $\log(G-1)$ are 
\begin{gather}
-\frac{\left(G-1\right)\left(1-T\right)}{2\left(1+GT\right)}+\frac{\left(G-1\right)\left(1+T\right)}{2\left(1-GT\right)}\\
= \frac{\left(G^2-1\right)T}{1-G^2T^2}\ ,
\end{gather}
the terms proportional to $\log(1+T)$ are
\begin{gather}
\frac{\left(G+1\right)\left(1+T\right)}{2\left(1+GT\right)}+\frac{\left(G-1\right)\left(1+T\right)}{2\left(1-GT\right)}\\
= \frac{\left(1-T^2\right)G}{1-G^2T^2}\ ,
\end{gather}
the terms proportional to $\log(1-T)$ are
\begin{gather}
-\frac{\left(G-1\right)\left(1-T\right)}{2\left(1+GT\right)}-\frac{\left(G+1\right)\left(1-T\right)}{2\left(1-GT\right)}\\
= -\frac{\left(1-T^2\right)G}{1-G^2T^2}\ ,
\end{gather}
the terms proportional to $\log(\frac{1}{2\left(1+GT\right)})=-\log(1+GT)-1$ are
\begin{gather}
\frac{\left(G+1\right)\left(1+T\right)}{2\left(1+GT\right)}-\frac{\left(G-1\right)\left(1-T\right)}{2\left(1+GT\right)}\\
= 1\ ,
\end{gather}
and finally the terms proportional to $\log(\frac{1}{2\left(1-GT\right)})=-\log(1-GT)-1$ are
\begin{gather}
-\frac{\left(G+1\right)\left(1-T\right)}{2\left(1-GT\right)}+\frac{\left(G-1\right)\left(1+T\right)}{2\left(1-GT\right)}\\
= -1\ .
\end{gather}

\noindent Collecting all these terms and the $\log(\frac{1+GT}{1-GT})$ term, equation \eqref{eq:sqeunfinished2} becomes

\begin{gather}
-\frac{\left(G^2-1\right)T}{1-G^2T^2}\log(G+1)+\frac{\left(G^2-1\right)T}{1-G^2T^2}\log(G-1)+\frac{\left(1-T^2\right)G}{1-G^2T^2}\log(1+T) \nonumber \\
-\frac{\left(1-T^2\right)G}{1-G^2T^2}\log(1-T)\underbrace{-\log(1+GT)-1+\log(1-GT)+1+\log(\frac{1+GT}{1-GT})}_{0}\\
= -\frac{\left(G^2-1\right)T}{1-G^2T^2}\left(\log(G+1)-\log(G-1)\right)+\frac{\left(1-T^2\right)G}{1-G^2T^2}\left(\log(1+T)-\log(1-T)\right)\\
= \frac{\left(1-T^2\right)G\log(\frac{1+T}{1-T})-\left(G^2-1\right)T\log(\frac{G+1}{G-1})}{1-G^2T^2}\ ,
\label{eq:sqeunfinished3}
\end{gather}
where in the first equality we regrouped terms and used the fact that the sum of the last five terms equals zero. The second equality follows from rewriting the logarithm terms.\\
\noindent Setting $G = \frac{1}{T}$, the denominator of equation \eqref{eq:sqeunfinished3} becomes zero. Luckily, the numerator $\left(1-T^2\right)\frac{1}{T}\log(\frac{1+T}{1-T})-\left(\frac{1}{T^2}-1\right)T\log(\frac{\frac{1}{T}+1}{\frac{1}{T}-1}) = \left(\frac{1}{T}-T\right)\log(\frac{1+T}{1-T})-\left(\frac{1}{T}-T\right)\log(\frac{1+T}{1-T}) = 0$, also becomes zero, implying that we can use L'H\^opital's rule to retrieve the limit. Differentiating the numerator from equation \eqref{eq:sqeunfinished3} with respect to $G$ gives

\begin{gather}
\left(1-T^2\right)\log(\frac{1+T}{1-T})+\frac{2T}{\ln2}-2GT\log(\frac{G+1}{G-1})\ ,
\label{eq:numerator}
\end{gather}
while differentiating the denominator from equation \eqref{eq:sqeunfinished3} gives

\begin{gather}
-2GT^2\ .
\label{eq:denominator}
\end{gather}
so that the quotient of equation \eqref{eq:numerator} and \eqref{eq:denominator} gives

\begin{gather}
\frac{-\left(1-T^2\right)\log(\frac{1+T}{1-T})-\frac{2T}{\ln2}+2GT\log(\frac{G+1}{G-1})}{2GT^2}\ .
\end{gather}

Setting $G = \frac{1}{T}$ we retrieve that

\begin{gather}
\lim_{G\rightarrow \frac{1}{T}} \frac{\left(1-T^2\right)G\log(\frac{1+T}{1-T})-\left(G^2-1\right)T\log(\frac{G+1}{G-1})}{1-G^2T^2}\\
= \lim_{G\rightarrow \frac{1}{T}} \frac{-\left(1-T^2\right)\log(\frac{1+T}{1-T})-\frac{2T}{\ln2}+2GT\log(\frac{G+1}{G-1})}{2GT^2}\\
= \frac{\left(T^2-1\right)\log(\frac{1+T}{1-T})-\frac{2T}{\ln2}+2\log(\frac{1+T}{1-T})}{2T}\\
= \frac{T^2+1}{2T}\log(\frac{1+T}{1-T})-\frac{1}{\ln2} \ .
\end{gather}

\noindent We see that for all three regimes ($G = \frac{1}{T},~G>\frac{1}{T}$ and $G<\frac{1}{T}$) equation \eqref{eq:finiten}, yields equation \eqref{eq:sqeunfinished3} in the asymptotic limit of $N\rightarrow \infty$. From this we retrieve our claim that 
\begin{align}
Q_2(\mathcal{N}_{\mathrm{PI}}),P_2(\mathcal{N}_{\mathrm{PI}}) \leq \Esq(\mathcal{N}_{\mathrm{PI}}) & \leq \frac{H(B|E_1'E_2')+H(B|F_1'F_2')}{2} \\
&=\frac{\left(1-T^2\right)G\log(\frac{1+T}{1-T})-\left(G^2-1\right)T\log(\frac{G+1}{G-1})}{1-G^2T^2}\ .
\label{eq:infintenfinal}
\end{align}

	\end{document}